\documentclass[prd, preprint, nofootinbib]{revtex4}%

\usepackage{amsfonts}
\usepackage{amsmath}
\usepackage{amssymb}
\usepackage{graphicx}
\usepackage{appendix}
\usepackage{hyperref}
\usepackage{color}

\newcommand{\dd}{{\rm{d}}} 

\newcommand{\im}{\mathrm{i}}

\begin{document}


\begin{flushright}
\ \vspace{-5.5mm}
$LIFT$--7-2.24
\end{flushright}
\vspace{-4.5mm}


\title{\LARGE
Black holes of type D revisited:\\ relating their various metric forms\\[1mm]}

\author{Hryhorii Ovcharenko}
\email{gregor\_ovcharenko@outlook.com, hryhorii.ovcharenko@matfyz.cuni.cz}
\affiliation{Charles University, Faculty of Mathematics and Physics,
Institute of Theoretical Physics,
V~Hole\v{s}ovi\v{c}k\'ach 2, 18000 Prague 8, Czechia}

\author{Ji\v{r}\'i Podolsk\'{y}}
\email{jiri.podolsky@mff.cuni.cz}  
\affiliation{Charles University, Faculty of Mathematics and Physics,
Institute of Theoretical Physics,
V~Hole\v{s}ovi\v{c}k\'ach 2, 18000 Prague 8, Czechia}

\author{Marco Astorino}
\email{marco.astorino@gmail.com}
\affiliation{Laboratorio Italiano di Fisica Teoretica (LIFT), \\
Via Archimede 20, I-20129 Milano, Italy}

\begin{abstract}
We investigate a complete family of spacetimes which represent black holes with (Kerr) rotation, Newman-Unti-Tamburino (NUT) twist, acceleration, electric and magnetic charges. These are exact solutions of the Einstein-Maxwell equations with any cosmological constant, such that the (non-null) electromagnetic field is aligned with both the double-degenerate principal null directions of the Weyl tensor. In particular, we explicitly relate various coordinates and the corresponding physical parameters of such solutions, namely the original Pleba\'nski-Demia\'nski (PD) form, the convenient Astorino (A) form which was found recently, and formally improved here (A$^+$), the Griffiths-Podolsk\'y (GP), and Podolsk\'y-Vr\'atn\'y (PV) form of the metric. It is demonstrated that, if properly mapped and physically interpreted, all these representations cover the complete class of type D black holes. Using the new A-parameters, the two main PD quartic metric functions are factorized into the product of quadratic expressions, enabling thus an explicit analysis. Moreover, we clarify the role of the twist parameter $\omega$, related to both the Kerr-like rotation and the NUT parameters $a$ and $l$, respectively. Special attention is payed to the elusive subclass of accelerating NUT black holes with ${a=0}$.
\end{abstract}

\date{\today}


\maketitle

\tableofcontents
\newpage

\section{Introduction}

The aim of this article is to elucidate mutual relations between various metric representations of a large family of solutions in the Einstein-Maxwell theory (with or without the cosmological constant) which belong to the type D of the Petrov-Penrose classification. In particular, we focus on stationary and axisymmetric solutions (i.e., with a couple of commuting Killing vectors $\partial_t$ and $\partial_\varphi$) such that two expanding repeated principal null directions (PNDs) of the Weyl tensor are both aligned with the two principal null directions of the electromagnetic field.\\

Such a family of spacetimes is relevant because it describes the most renowned black holes in general relativity, starting from the static and spherically symmetric metrics, such as the Schwarzschild line element, the C-metric describing accelerating black holes, the Newman-Unti-Tamburino (NUT) twisting spacetime, to a general stationary rotating, accelerating, and charged Kerr-Newman solution \cite{Stephanietal:2003, GriffithsPodolsky:2009}. This class is often identified with the Pleba\'nski-Demia\'nski family \cite{PlebanskiDemianski:1976} (but see also the earlier work of Debever \cite{Debever:1971}), subsequently investigated in detail in \cite{GriffithsPodolsky:2005, GriffithsPodolsky:2006, PodolskyGriffiths:2006, PodolskyVratny:2021, PodolskyVratny:2023}.\footnote{Actually, this is the subclass of all type D spacetimes for the theory under consideration. Other solutions, including non-expanding cases, were studied in a number of works. See, e.g., the review \cite{VandenBergh:2017}.}

Recently a large class of such type~D solutions was systematically investigated by means of the solution generating technique, and a nice new metric form was thus obtained \cite{Astorino:2024b, Astorino:2024a}. This novel spacetime representation has the advantage to directly contain the limits to \emph{all} the subcases of type~D black holes contained in the general Pleba\'nski-Demia\'nski solution, including also the peculiar accelerating solutions with (just) the NUT parameter, which  was previously considered to exist only \emph{outside} the type D class \cite{PodolskyVratny:2020}, \cite{PodolskyVratny:2021}. It came as a surprise because the only accelerating black holes with NUT parameter known before \cite{Astorino:2024a}, namely the Chng-Mann-Stelea metric \cite{ChngMannStelea:2006} investigated in \cite{PodolskyVratny:2020}, were of a general algebraic type~I (see \cite{Astorino:2023elf, Astorino:2023b} for the rotating and charged generalization).

\newpage

This discovery of a novel general form of type D metric, which comprises accelerating black holes with NUT parameter, naturally opens the way to questions about the actual generality of the Pleba\'nski-Demia\'nski metric and its different parameterizations, namely:
\begin{enumerate}
\item Is the Pleba\'nski-Demia\'nski solution the most general black hole spacetime of type D, or is the metric presented in \cite{Astorino:2024b} its extension?
\item Might the metric of \cite{Astorino:2024b} be just another equivalent reparametrization of the Pleba\'nski-Demia\'nski metric, but more suitable for description of all type D black hole specializations?
\item What is the relation between the new spacetime of \cite{Astorino:2024b} and the type D metrics known so far in the literature, such as those in \cite{GriffithsPodolsky:2005, GriffithsPodolsky:2006, PodolskyGriffiths:2006, PodolskyVratny:2021, PodolskyVratny:2023}?
\item Why the accelerating NUT black holes have not been explicitly identified in previous works \cite{GriffithsPodolsky:2005, GriffithsPodolsky:2006}?
\item Which is the more appropriate/convenient parametrization to describe the physical and geometrical properties of the whole class of accelerating Kerr-Newman-NUT black holes?
\end{enumerate}

\vspace{2mm}

In our paper we address these open questions. In Section~\ref{sc:Astorino-form} we start by revisiting the solution of \cite{Astorino:2024b}, denoted here as A, putting it into a simpler metric form which we will denote A$^+$. In subsequent Section~\ref{sc:PD-form} the transformation from the Astorino metric to the original Pleba\'nski-Demia\'nski coordinates, together with explicit relation of the physical A and A$^+$ parameters to the PD integration constants, is presented (full details can be found in Appendix~\ref{systematic-derivation}). This leads to a factorized form of the PD metric functions, and their simplification for various special cases. Transformation to the Griffiths-Podolsk\'y form of this family of black-hole spacetimes is presented in Section~\ref{sc:GP-form}, and their special cases are discussed in Section~\ref{sc:the-special-cases}, after elucidating the role of the twist parameter~$\omega$ and clarifying the physical dimensionality of the parameters. Relation to the Podolsk\'y-Vratn\'y metric representation is contained in Section~\ref{sc:PV-form}. Section~\ref{sc:special-cases} summarizes and compares the key special cases in A, PD, GP, and PV metric forms, followed by concluding remarks in Section~\ref{sc:conclusions}.\\

For convenience of the reader, we summarize our nomenclature and conventions for A, PD, GP, and PV coordinates and parameters in Table~\ref{Tab-summary-of-metrics}, together with references to original articles. These metrics admit any value of the cosmological constant $\Lambda$, but in this paper we only investigate black hole spacetimes with ${\Lambda=0}$.

\vspace{10mm}

\begin{table}[!h]
\begin{center}
\caption{\label{Tab-summary-of-metrics} Summary of the metrics for type~D black-hole spacetimes studied in this paper.}
\vspace{2.0mm}
\begin{tabular}{l|c|c|c|c}
\hline
\hline
metric form\  & \ eq.\  & \ original source\  &  \ coordinates\ & \  parameters\ \\[2pt]
\hline
\hline
A             & \eqref{init_metr}   & Astorino \cite{Astorino:2024b} & ${(r, x, t, \varphi)}$ &
   ${\alpha, a, l, m, e, g, \Lambda; \,C_f }$ \\
A$^+$          & \eqref{ds2_simpl}   & this paper &  \\
\hline
PD            & \eqref{oldPDMetric} & Pleba\'nski-Demia\'nski \cite{PlebanskiDemianski:1976}
   & ${(r', p', \tau', \sigma')}  $ &
   ${k', n', \epsilon', m', e', g', \Lambda}$ \\
PD$_{\alpha}$ & \eqref{PD-GP-form}  & \cite{GriffithsPodolsky:2005, GriffithsPodolsky:2006}
 & ${(r', x', \tau', \phi')}$ &
   ${\alpha', k', n', \epsilon', m', e', g', \Lambda; \,c }$ \\
PD$_{\alpha\omega}$&\eqref{PleDemMetric-again} &  \cite{GriffithsPodolsky:2005, GriffithsPodolsky:2006}
 & ${(r', x', \tau', \phi')}$ &
   ${\alpha', k', n', \epsilon', m', e', g', \Lambda; \,\omega, c }$ \\
\hline
GP$_{\omega}$      & \eqref{ds2_accel_kerr_new_polar} & Griffiths-Podolsk\'y \cite{GriffithsPodolsky:2005, GriffithsPodolsky:2006, PodolskyGriffiths:2006}
 & ${(\tilde{r}, \theta, t, \varphi)}$ &
   ${\tilde{\alpha}, \tilde{a}, \tilde{l}, \tilde{m}, \tilde{e}, \tilde{g}, \Lambda; \,\omega, c }$ \\
PV$_{\omega}$            & \eqref{PV-metric} & Podolsk\'y-Vr\'atn\'y \cite{PodolskyVratny:2021, PodolskyVratny:2023}
 & ${(\hat{r}, \theta, t, \varphi)}  $ &
   ${\hat{\alpha}, \hat{a}, \hat{l}, \hat{m}, \hat{e}, \hat{g}, \Lambda; \,\omega, c }$  \\[1pt]
\hline
\hline
\end{tabular}
\end{center}
\end{table}

\vspace{10mm}

\begin{figure}[!h]
\begin{center}
\includegraphics[scale=0.88]{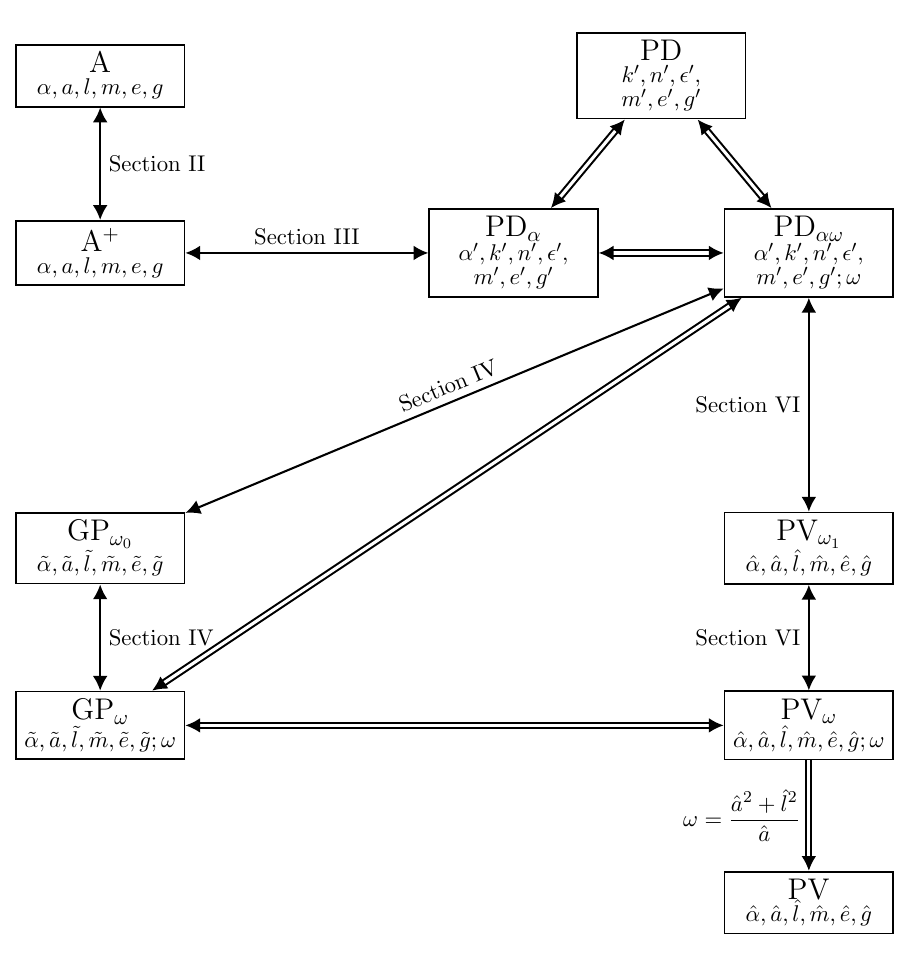}

\vspace{-2.0mm}
\caption{
Scheme of the relations between different metric forms of black-hole solutions studied in this paper for the case ${\Lambda=0}$. Here, A stands for the original Astorino solution, while A$^+$ denotes a new compact form of the same solution introduced in Section~\ref{sc:Astorino-form}. PD is the original Pleba\'nski-Demia\'nski solution, $\mathrm{PD_{\alpha}}$ is the Pleba\'nski-Demia\'nski solution with acceleration~$\alpha$, and $\mathrm{PD}_{\alpha \omega}$ with with both the acceleration~$\alpha$ and the twist parameter~$\omega$. $\mathrm{GP_{\omega_0}}$ stands for the Griffiths-Podolsk\'y form with ${\omega=\omega_0}$, and $\mathrm{GP_{\omega}}$ is the same metric but with an arbitrary value of~$\omega$, while $\mathrm{PV_{\omega_1}}$ is the Podolsk\'y-Vr\'atn\'y form with ${\omega=\omega_1}$, and $\mathrm{PV_{\omega}}$ stands for the same form but with an arbitrary~$\omega$. Finally, PV denotes the same form, but with the special choice ${\omega=\dfrac{\hat{a}^2+\hat{l}^2}{\hat{a}}}$.}
\label{Fig-scheme}
\end{center}
\end{figure}

Mutual relations between all these metric forms are shown in the scheme on Figure~\ref{Fig-scheme}. Particular connections are presented in full detail in Sections which are indicated by the corresponding arrows between them. Namely, the double-arrows show equivalence proven already in previous works, the single-arrows are the equivalences proven in this paper.

\newpage

\newpage

\section{The Astorino most general metric form}
\label{sc:Astorino-form}

In 2024, Astorino \cite{Astorino:2024b} presented a very convenient explicit metric for the most general type~D black hole (with a cosmological constant $\Lambda$ and doubly-aligned electromagnetic field) in the form\footnote{We have relabeled the parameters, namely $\alpha$ to $-\alpha$, $l$ to $-l$, and $p$ to $g$.}
\begin{equation}
\dd s^{2}=-f(r,x)\,\big[\dd t-\omega(r,x)\,\dd\varphi\big]^{2}+\frac{1}{f(r,x)}\bigg[
e^{2\gamma(r,x)}\left(  \frac{\dd r^{2}}{\Delta_{r}(r)}+\frac{\dd x^{2}}{\Delta_{x}(x)}\right)
+ \varrho^{2}(r,x)\,\dd\varphi^{2}\bigg],\label{init_metr}
\end{equation}
where the metric functions are
\begin{align}
f=&\ \frac{[1+\alpha^{2}(l^{2}-a^{2})x^{2}]^{2}\,\Delta_{r}
-[a+2\alpha l\,r+\alpha^{2}a\,r^{2}]^2\,\Delta_{x}}
{(1-\alpha r x)^2\rho^2}, \label{f}\\[1mm]
\omega=&\ \frac{(a+2l\,x+a\,x^{2})[1+\alpha^{2}(l^{2}-a^{2})x^{2}]\,\Delta_{r}%
+(r^{2}+l^{2}-a^{2})(a+2\alpha l\,r+\alpha^{2}a\,r^{2})\,\Delta_{x}}
{[1+\alpha^{2}(l^{2}-a^{2})x^{2}]^{2}\,\Delta_{r}-[a+2\alpha l\,r+\alpha^{2}a\,r^{2}%
]^{2}\,\Delta_{x}},\\[1mm]
e^{2\gamma}=&\ C_f\,\dfrac{[1+\alpha^2(l^2-a^2)x^2]^2\,\Delta_r-(a+2\alpha l\, r+\alpha^2 a\, r^2)^2\,\Delta_x}{(1-\alpha r x)^4},\\[1mm]
\varrho^2=&\ \dfrac{\Delta_r\,\Delta_x}{(1-\alpha r x)^4},\\[1mm]
\Delta_r=&\ (1-\alpha^2r^2)\big[(r-m)^2 - (m^2+l^2-a^2-e^2-g^2)\big] \nonumber\\
   & \hspace{20mm} - \frac{\Lambda}{3}\Big(\,\frac{3l^2}{1+\alpha^2 a^2}\,r^2
     + \frac{4\alpha a l}{1+\alpha^2 a^2}\,r^3 + r^4 \Big) ,\label{delta_r_init}\\[1mm]
\Delta_x=&\ (1-x^2)\big[(1-\alpha m \,x)^2-\alpha^2x^2(m^2+l^2-a^2-e^2-g^2)\big] \nonumber\\
   & \hspace{20mm} - \frac{\Lambda}{3}\Big(\,\frac{3l^2}{1+\alpha^2 a^2}\,x^2
    +\frac{4al}{1+\alpha^2 a^2}\,x^3 + \frac{a^2+\alpha^2(a^2-l^2)^2}{1+\alpha^2 a^2}\,x^4 \Big) , \label{delta_x_init}
\end{align}
and $\rho^2$ in the numerator of (\ref{f}) is defined as
\begin{equation}
\rho^2= (l+a\,x)^2 + 2\alpha\,l\,(a+2l\,x+a\,x^2)\,r + \alpha^2(a^2-l^2)^2\,x^2 + [1+\alpha^2(a+l\,x)^2\,]\,r^2 . \label{rho2expl}
\end{equation}

It was argued in \cite{Astorino:2024b} that $m$ is the mass parameter, $a$ denotes rotation, $l$ in the NUT parameter, $\alpha$ is acceleration, $e$ and $g$ are electric and magnetic charges,\footnote{More precisely, they are directly related to physical conserved charges only in some special cases, not in the general case.} and $C_f$ is an additional normalization constant (which could be related to conicity).\footnote{Here, without loss of generality, we have set the auxiliary constant parameters $\epsilon, \kappa, \xi, \chi$ which appear in Eq.~(2.7) in \cite{Astorino:2024b} to zero. See also the Wolfram Mathematica Notebook in Supplementary Material of \cite{Astorino:2024b}.} Here  we naturally assume that $m, a, \alpha$ are positive (or zero), while $l, e, g, \Lambda$ can take any value.

We should also clarify that these quantities have \emph{usual physical dimensions}, namely $m, a, l, e, g$ have the dimension of length, $\alpha$ and $\sqrt{\Lambda}$ are inverse length, and $C_f$ is dimensionless. The coordinates $t, r$ have dimension of length while $x, \varphi$ are dimensionless. Consequently, $\Delta_x, f$ are dimensionless, $ \Delta_r, \varrho^2, \rho^2, e^{2\gamma}$  have the dimension of length squared, and $\omega(r,x)$ has the dimension of length.\\

The electromagnetic field is given by the vector potential ${A(r,x)=A_t\,\dd t + A_\varphi\,\dd\varphi}$ with
\begin{align}
 A_t =& \,\sqrt{\frac{1+\alpha^2(a^2-l^2)}{1+\alpha^2a^2}}\,\frac{-1}{\rho^2\, l}
       \Big[ \big[\,g (r + \alpha a l +  \alpha^2 a^2 r) + e l \,\big]\,r  \nonumber\\
 & \hspace{3mm} + l\,(g-\alpha a e) (a+2 \alpha l\, r+\alpha^2a\,r^2)\,x
  \label{At}\\
 & \hspace{3mm} +(a + \alpha l\,r)\big[\,ag + \alpha l(gr-el) + \alpha^2a\,[g(a^2-l^2)-el\,r] \big]\,x^2\Big]
   + \frac{g}{l} \,,  \nonumber \\
 A_\varphi =& \,\sqrt{\frac{1+\alpha^2(a^2-l^2)}{1+\alpha^2a^2}}\,\frac{x}{\rho^2}
         \Big[
    \alpha l^3 (e+\alpha ag)x  + l^2\big[\alpha^2a(e+\alpha a g)\,xr + (\alpha a e-g)\big] \label{Aphi}\\
 & \hspace{3mm}+ (1+\alpha^2a^2)\big[ ae(x+\alpha\,r)\,r - gal\,x + g(r-\alpha a^2x)\,r + el(2+\alpha xr)r \big]  \Big] - (a+\omega_c)\, A_t\,,  \nonumber
\end{align}
where $\omega_c$ is a constant (for the choice ${\omega_c=-a}$ the last term vanishes).
The apparently problematic terms ${l^{-1}}$ are included to remove the divergencies in the limit ${l \to 0}$.

Now, to simplify the metric (\ref{init_metr})--(\ref{delta_x_init}), it is useful to define the following functions
\begin{align}
    A(x) &:= 1 + \alpha^2(l^2-a^2)\,x^2,\label{Aw}\\
    B(x) &:= a + 2l\,x + a\,x^2,\label{Bw}\\
    C(r) &:= a + 2\alpha l\,r + \alpha^2 a\, r^2,\label{Cw}\\
    D(r) &:= (l^2-a^2) + r^2,\label{Bu}
\end{align}
and
\begin{align}
    \Omega(r,x) &:= 1-\alpha\, r\, x .\label{Om_cf}
\end{align}
Interestingly, their specific combination,
\begin{align}
\rho^2 =  AD+BC, \label{rho2def}
\end{align}
gives exactly the complicated expression introduced in (\ref{rho2expl}). Notice also that \eqref{Aw}--\eqref{Bu} are just quadratic polynomials, and \eqref{rho2expl} also includes up to the second power, both in $x$ and $r$.

With these shorthands, we can rewrite the metric functions $f$,~$\omega$ and~$\gamma$ as
\begin{align}
    f=&\ \dfrac{A^2\Delta_r-C^2\Delta_{x}}{\Omega^2\rho^2},\label{f_siml} \\
    \omega=&\ \dfrac{ AB\,\Delta_r + CD\,\Delta_x}{A^2\,\Delta_r-C^2\,\Delta_x},\label{om_simpl} \\
    e^{2\gamma}=&\ C_f\,\dfrac{A^2\,\Delta_r - C^2\,\Delta_x}{\Omega^4},\label{gamma_simpl} \\
    \varrho^2=&\ \dfrac{\Delta_r\,\Delta_x}{\Omega^4}, \label{rho_simpl}
\end{align}
and after substituting (\ref{f_siml})--(\ref{rho_simpl}) into (\ref{init_metr}) we get the metric
\begin{align}
    \dd s^2 =&\ \dfrac{1}{\Omega^2}\bigg[-\dfrac{A^2\Delta_r-C^2\Delta_x}{\rho^2}
    \Big(\dd t-\dfrac{AB\,\Delta_r + CD\,\Delta_x}{A^2\Delta_r-C^2\Delta_x}\,\dd\varphi\Big)^2
    \nonumber\\
    &\qquad +\dfrac{\rho^2\Delta_r\,\Delta_x}{A^2\Delta_r-C^2\Delta_x}\,\dd\varphi^2
    + C_f\,\rho^2 \Big(\,\dfrac{\dd r^2}{\Delta_r} + \dfrac{\dd x^2}{\Delta_x}\,\Big)\bigg].
\end{align}

Next, we simplify the parts depending on $\dd t$ and $\dd \varphi$, namely
\begin{align}
&-\dfrac{A^2\Delta_r-C^2\Delta_x}{\rho^2}\,\dd t^2
 + 2\,\dfrac{AB\,\Delta_r +CD\,\Delta_x}{\rho^2}\,\dd t \,\dd\varphi\nonumber\\
&-\dfrac{(AB\,\Delta_r + CD\,\Delta_x)^2}{\rho^2(A^2\Delta_r-C^2\Delta_x)}\,\dd\varphi^2
+\dfrac{\rho^2\Delta_r\,\Delta_x}{A^2\Delta_r-C^2\Delta_x}\,\dd\varphi^2.\label{dtphi2}
\end{align}
Using the definition (\ref{rho2def}) the last two terms combine to
\begin{align}
   \dfrac{D^2\Delta_x - B^2\Delta_r}{\rho^2}\,\dd\varphi^2,
\end{align}
so that the whole complicated expression (\ref{dtphi2}) ``miraculously'' simplifies to
\begin{align}
   -\dfrac{\Delta_r}{\rho^2}(A\,\dd t-B\,\dd\varphi)^2+\dfrac{\Delta_x}{\rho^2}(C\,\dd t+D\,\dd\varphi)^2.
\end{align}

This allows us to finally write the Astorino \emph{complete metric} \emph{in a very compact and explicit form} as
\begin{equation}
    \dd s^2=\dfrac{1}{\Omega^2}\bigg[-\dfrac{\Delta_r}{\rho^2}(A\,\dd t - B\,\dd\varphi)^2
    + \dfrac{\Delta_x}{\rho^2}(C\,\dd t + D\,\dd\varphi)^2
    + C_f\,\rho^2 \Big(\,\dfrac{\dd r^2}{\Delta_r} + \dfrac{\dd x^2}{\Delta_x}\,\Big)\bigg]
    \label{ds2_simpl}.
\end{equation}
It may naturally be given a \emph{nickname ``A$^+$ metric''}.
Recall that $\rho^2$ is the polynomial expression \eqref{rho2expl}, the quadratic functions ${A, B, C, D, \Omega}$ have the form \eqref{Aw}--\eqref{Om_cf}, and the two quartics ${\Delta_r, \Delta_x}$  are given by \eqref{delta_r_init}, \eqref{delta_x_init}.

The Astorino (A) solution \eqref{init_metr} rewritten in the new compact (A$^+$) form of the metric  \eqref{ds2_simpl} resembles the Griffiths-Podolsk\'y (GP) representation \cite{GriffithsPodolsky:2009, GriffithsPodolsky:2005, GriffithsPodolsky:2006, PodolskyGriffiths:2006, PodolskyVratny:2021, PodolskyVratny:2023} of the family of type D black holes in the Pleba\'nski-Demia\'nski (PD) class of electrovacuum solutions with~$\Lambda$. An exact relation between these metric forms, and the physical parameters of the black holes, will be derived and investigated later in Sections~\ref{sc:GP-form}--\ref{sc:PV-form}.

However, first we will present an explicit and direct transformation from the Astorino metric \eqref{init_metr}, that is \eqref{ds2_simpl}, to the original Pleba\'nski-Demia\'nski form of the metric. This will demonstrate, that \emph{both} the Astorino class of solutions \emph{and} the Pleba\'nski-Demia\'nski class of solutions are equivalent, and that represent \emph{all} solutions of a given type, including the elusive accelerating (purely) NUT black holes of algebraic type~D.

\newpage

\section{Transformation to the Pleba\'nski-Demia\'nski metric form}
\label{sc:PD-form}

A general Pleba\'nski-Demia\'nski metric representing all solutions of \hbox{Einstein-Maxwell-$\Lambda$} equations of algebraic type~D (with double aligned, non-null electromagnetic field) --- which includes black holes of this type --- is originally given by Eq.~(3.30) in the seminal paper~\cite{PlebanskiDemianski:1976}.
It is also repeated (with a slight modification of the symbols used) in Chapter~16 of~\cite{GriffithsPodolsky:2009} as Eqs.~(16.1) and $(16.2)$, namely
  \begin{equation}
  \begin{array}{l}
 {\displaystyle \dd s^2=\frac{1}{(1-p'\,r')^2} \Bigg[
 -\frac{Q'\,(\dd\tau'-p'^{\,2}\,\dd\sigma')^2}{r'^{\,2}+p'^{\,2}}
 +\frac{P'\,(\dd\tau'+r'^{\,2}\,\dd\sigma')^2}{r'^{\,2}+p'^{\,2}} } \\[10pt]
 \hskip12pc {\displaystyle +\frac{r'^{\,2}+p'^{\,2}}{{P'}}\,\dd p'^{\,2}
 +\frac{r'^{\,2}+p'^{\,2}}{Q'}\,\dd r'^{\,2} \Bigg]} \,.
 \end{array}
  \label{oldPDMetric}
  \end{equation}
This contains two quartic functions
  \begin{equation}
  \begin{array}{l}
  {P'}(p') = k' +2 n'p' - \epsilon'p'^{\,2} +2 m'p'^{\,3}-(k'+e'^2+g'^2+\Lambda/3)\,p'^{\,4} \,, \\[8pt]
  {Q'}(r') =(k'+e'^2+g'^2) -2m'r' +\epsilon'r'^{\,2} -2n'r'^{\,3}-(k'+\Lambda/3)\,r'^{\,4} \,,
  \end{array}
 \label{oldPQeqns}
  \end{equation}
with 7 arbitrary real parameters $\Lambda$, $e'$, $g'$, $m'$, $n'$, $\epsilon'$, $k'$ (the parameter $\gamma$ of~\cite{PlebanskiDemianski:1976} is obtained by putting ${k'=\gamma-g'^2-\Lambda/6}$). Here $\Lambda$ is the cosmological constant, while~$e'$ and~$g'$ represent electric and magnetic charges, as the vector potential reads
\begin{align}
A = - \frac{e' + {\rm i}\,g'}{r' + {\rm i}\,p'}\,\big(\,\dd\tau'  -  {\rm i}\,p'\,r'\, \dd\sigma' \,\big).\label{A_PD}
\end{align}

In~\cite{GriffithsPodolsky:2005} a convenient rescaling of the original PD metric \eqref{oldPDMetric}, \eqref{oldPQeqns} was performed, namely
${p' \mapsto\sqrt{\alpha\omega}\,p'}$, ${r' \mapsto \sqrt{\alpha/\omega}\,r'}$, ${\sigma' \mapsto \sqrt{\omega/\alpha^3}\,\sigma'}$, ${\tau' \mapsto =\sqrt{\omega/\alpha}\,\tau'}$,
with the relabelling of constants
${m'+\im\,n' \mapsto ({\alpha'/\omega})^{3/2}(m'+\im\, n')}$,
${e'+\im\,g' \mapsto ({\alpha'/\omega})(e'+\im\, g')}$,
${\epsilon' \mapsto ({\alpha'/\omega})\,\epsilon'}$,
${k' \mapsto \alpha'^2k}$.
This introduced two important kinematic parameters $\alpha$ and $\omega$, later interpreted as the \emph{acceleration} and the \emph{twist} of the black hole, respectively. Such a rescaled metric, we will denote as PD$_{\alpha\omega}$, reads
  \begin{eqnarray}
  &&\hskip-3pc\dd s^2=\frac{1}{(1-\alpha'\, p'\,r')^2} \Bigg[
 -\frac{{\cal Q}}{r'^{\,2}+\omega^2p'^{\,2}}\,(\dd\tau'-\omega p'^{\,2}\dd\sigma')^2
 +\frac{{\cal P}}{r'^{\,2}+\omega^2p'^{\,2}}\,(\omega\dd\tau'+r'^{\,2}\dd\sigma')^2 \nonumber \\
  &&\hskip12pc
  +\frac{r'^{\,2}+\omega^2p'^{\,2}}{{\cal Q}}\,\dd r'^{\,2}
  +\frac{r'^{\,2}+\omega^2p'^{\,2}}{{\cal P}}\,\dd p'^{\,2} \Bigg], \label{PleDemMetric}
  \end{eqnarray}
  where the key functions are
 \begin{equation}
 \begin{array}{l}
{\cal P}(p') = k' +2\omega^{-1}n'p' -\epsilon'p'^{\,2} +2\alpha' m'p'^{\,3}
-\big[\alpha'^{\,2}(\omega^2 k'+e'^{\,2}+g'^{\,2})+\omega^2\Lambda/3\big]\,p'^{\,4} \,,\\[8pt]
{\cal Q}(r') = (\omega^2 k'+e'^{\,2}+g'^{\,2}) -2m'r' +\epsilon' r'^{\,2} -2\alpha'\omega^{-1}n'r'^{\,3}
-(\alpha'^{\,2}k'+\Lambda/3)\,r'^{\,4} \,,
 \end{array}
 \label{PQeqns}
 \end{equation}
see Eqs.~(16.5) and $(16.6)$ in~\cite{GriffithsPodolsky:2009}, with the vector potential
\begin{align}
A = - \frac{e' + {\rm i}\,g' }{r' + {\rm i}\,\omega\,p'}\,\big(\,\dd\tau'  -  {\rm i}\,p'r' \dd\sigma' \,\big).\label{A_PD-scaled}
\end{align}

Here we considered the Pleba\'nski-Demia\'nski metric, denoted as PD$_{\alpha}$,
\begin{align}
    \dd s^2=\dfrac{1}{{(1-\alpha'\,r'\,x')}^{\,2}}\bigg[
    &- \dfrac{Q'}{{r'}^{\,2}+{x'}^{\,2}}\,(\dd\tau'  -  {x'}^{\,2}\, \dd\phi' )^2
     + \dfrac{P'}{{r'}^{\,2}+{x'}^{\,2}}\,(\dd\tau'  +  {r'}^{\,2}\, \dd\phi' )^2 \nonumber\\
    &+ c^2 ({r'}^{\,2}+{x'}^{\,2}) \Big(\,\dfrac{\dd {r'}^{\,2}}{Q'} + \dfrac{\dd {x'}^{\,2}}{P'}\,\Big)\bigg], \label{PD-GP-form}
\end{align}
with the metric functions
 \begin{equation}
 \begin{array}{l}
P'(x') = k'+2n' x' - \epsilon'x'^{\,2} +2\alpha'm' x'^{\,3}
- \big[\alpha'^{\,2} ({k'}+{e'}^2 + {g'}^2)+c^2\Lambda/3\big]\,x'^{\,4}   \,, \\[8pt]
Q'(r') = (k'+e'^2 + g'^2)-2m' r' + \epsilon'r'^{\,2} - 2\alpha' n' \,r'^{\,3} - (\alpha'^{\,2} k' + c^2\Lambda/3)\,r'^{\,4} \,.
 \end{array}
 \label{P'Q'eqns}
 \end{equation}
Changing $x'$ to $p'$, $\phi'$ to $\sigma$, and setting ${c=1}$, these are exactly the expressions \eqref{PQeqns} for the twist parameter ${\omega=1}$. They are equivalent to the metric functions~$(16.6)$ in~\cite{GriffithsPodolsky:2009}.\\

From now on, we will \emph{only consider the case} ${\Lambda=0}$. Generalization to any value of the cosmological constant will be presented in our subsequent paper elsewhere.\\

A \emph{direct transformation} of coordinates between the original Astorino metric \eqref{init_metr}--\eqref{delta_x_init}, equivalent to the A$^+$ metric \eqref{ds2_simpl}, and the Pleba\'nski-Demia\'nski metric \eqref{PD-GP-form}, \eqref{P'Q'eqns} is
\begin{align}
t &=\ \dfrac{a}{\alpha(a^2-l^2)}\,\sqrt{\dfrac{K-1}{\sqrt{I^2 \mp J^2}}}\,
    \Big[\big[K-\alpha^2(a^2-l^2)\big]\,\tau' + \frac{K-1}{\alpha^2(a^2-l^2)}\,\phi'\,\Big],\label{direct_transformation_A-PD-t}\\
\varphi &=\ \dfrac{1}{\alpha(a^2-l^2)}\,\sqrt{\dfrac{K-1}{\sqrt{I^2 \mp J^2}}}\,
    \Big[\phi'-\alpha^2(a^2-l^2)\,\tau'\,\Big],\label{direct_transformation_A-PD-phi}\\
x         &=\ \frac{aK\,x'-l}{aK-l\alpha^2(a^2-l^2)\,x'}\,, \label{direct_transformation_A-PD-x}\\
\alpha\,r &=\ \frac{aK(a^2-l^2)\,\alpha\,r'-al(K-1)}{(a^2K-l^2) - l(a^2-l^2)\,\alpha\,r'}\,, \label{direct_transformation_A-PD-r}
\end{align}
in which the convenient (auxiliary) \emph{dimensionless constants} $I$, $J$, and $K$ are defined as
\begin{align} \label{defI-and -defJ}
    I := &\ 1+\alpha^2(a^2-l^2)\, , \nonumber\\
    J := &\ 2 \alpha\,\dfrac{l}{a}\,\sqrt{|a^2-l^2|}\,,\\
    2K := &\ I + \sqrt{I^2 \mp J^2}\,. \nonumber
\end{align}
Here the upper sign is used when ${a^2>l^2}$, while the lower sign is used in the complementary case ${a^2<l^2}$ (and ${a=0}$ in particular). Therefore,
\begin{align} \label{I2-pm-J2}
    a^2(I^2 \mp J^2) = a^2 [1+\alpha^2(a^2-l^2)]^2 + 4\alpha^2 l^2(l^2-a^2)\,.
\end{align}
and
\begin{align} \label{K-pm}
    2aK = a\,[1+\alpha^2(a^2-l^2)] + \sqrt{ a^2[1+\alpha^2(a^2-l^2)]^2 + 4\alpha^2 l^2(l^2-a^2)}\,,
\end{align}
which are useful explicit relations to be employed later.

It systematically follows from the procedure presented in Appendix~A, namely as a combination of expressions \eqref{def-x-and-r-1}, \eqref{def-x-and-r} with \eqref{def-x-and-r-with X0-and-R_0}, \eqref{transf-xi'}, \eqref{transf-chi'}, \eqref{choice-of-R0-E1}, \eqref{choice-of-R0-E}.

Actually, the transformation has a simple structure. For $t$ and $\varphi$ it is just a \emph{linear combination}, and for $x$ and $r$ it is a \emph{fraction} of \emph{linear expressions} of the respective coordinates (a real version of the M\"obius transformation). For ${l=0}$ this reduces to ${x=x'}$, ${r=a\,r'}$.

Let us observe that the coordinates $\tau', \phi'$ and the constant~$c$ in the metric \eqref{PD-GP-form} have the physical dimension of length, while $r', x'$ are dimensionless. Also the PD coefficients $\alpha', k', n', \epsilon', m', e', g'$, and thus both the metric functions $P', Q'$, are dimensionless.\footnote{There are other possibilities.
By a rescaling of the coordinates ${\tau' \mapsto c\,\tau'}$ and ${\phi' \mapsto c\,\phi'}$, the constant $c^2$ can be removed from the last two terms in \eqref{PD-GP-form} because it then becomes an overall constant conformal factor determining the specific physical scale of the metric in the square brackets with dimensionless coordinates $\tau', \phi', r', x'$. Alternatively, by performing ${r' \mapsto r'/c}$ and ${x' \mapsto x'/c}$ all the new coordinates $\tau', \phi', r', x'$ have the dimension of length (after the PD parameters are properly rescaled).}\\

As explained in full detail in Appendix, to exactly identify the metrics A and PD$_{\alpha}$, that is \eqref{init_metr} and \eqref{PD-GP-form}, it is also necessary to perform a \emph{constant dimensionless scaling} such that the conformal factor in \eqref{ds2-new-in-PD-form} is
\begin{equation}\label{conformal-relation-repeated}
\Omega'^{\,2} = S^2 \big(\,1 -\alpha'\,r'\,x'\,\big)^2,
\qquad
\hbox{where}
\qquad
S^2   = \frac{a^2}{\alpha^4\,|a^2-l^2|^3}\,\frac{K-1}{\sqrt{I^2 \mp J^2}},
\end{equation}
see \eqref{conformal-relation} and \eqref{S^2}. This specific rescaling is already included in the transformation of $t$ and $\varphi$ given by \eqref{direct_transformation_A-PD-t} and \eqref{direct_transformation_A-PD-phi}.\\

Even more importantly, the transformation \emph{uniquely relates the PD acceleration parameter} $\alpha'$ in  $\Omega'^{\,2}$ and the  more convenient acceleration parameter $\alpha$ as
\begin{align}\label{direct_transformation_A-PD-acceleraation}
\alpha' &=\ \alpha\,a\,\big[\,K - \alpha^2(a^2-l^2)\big],
\end{align}
see \eqref{alpha'} which follows from \eqref{c_0^2-alpha'}. Written explicitly it terms of the new Astorino parameters, it reads
\begin{align}\label{direct_transformation_A-PD-acceleration-explicit}
\alpha' &=\ \alpha\,\,\tfrac{1}{2}\big[\,a - \alpha^2 a (a^2-l^2)
+ \sqrt{a^2+\alpha^4 a^2(a^2-l^2)^2+2\alpha^2(a^2-l^2)(a^2-2l^2)}\, \big].
\end{align}
By setting ${\alpha=0}$ or ${l=0}$ we get $\alpha'=\alpha\,a$, while by setting ${a=0}$ we get ${\alpha'=\alpha^2 l^2}$.\\

Relations between the other six parameters in the Pleba\'nski-Demia\'nski metric functions  \eqref{P'Q'eqns}
are more involved, namely
\begin{align}\label{direct_transformation_A-PD_parameters-simplified}
k'  &=\ \frac{1}{I^2 \mp J^2}\,\,\frac{K-1}{\alpha^2(a^2-l^2)}\,\,I\, L\,, \nonumber\\
n'  &=\ \frac{-1}{I^2 \mp J^2}\,
      \Big[\,\frac{K-1}{\alpha^2(a^2-l^2)}\,\,I\, M  - \big[1-\alpha^2(a^2-l^2)\big]\,L\,\frac{l}{a}\, \Big], \nonumber\\
\epsilon'  &=\ \frac{1}{I^2 \mp J^2}\, \big[1-\alpha^2(a^2-l^2)\big]
      \Big( I\, L  + 4\,M\,\frac{l}{a} \Big)
      - (e^2+g^2)\frac{K-1}{(a^2-l^2)}\,\frac{I}{\sqrt{I^2 \mp J^2}} \,,\\
\alpha' m' &=\ \frac{1}{I^2 \mp J^2}\, \Big[ \,
       \big(K-1\big)\,I\, M
       \nonumber\\
&      \hspace{17mm}+ \big[1-\alpha^2(a^2-l^2)\big]\Big( I\, M
       + \alpha^2 \big[(a^2-l^2)\,L+ (e^2+g^2)\sqrt{I^2 \mp J^2}\,\big]\frac{l}{a} \Big)\Big], \nonumber\\
{e'}^2 + {g'}^2 &=\ ({e}^2 + {g}^2)\, \frac{K-1}{\alpha^2(a^2-l^2)^2}\,\frac{I}{\sqrt{I^2 \mp J^2}} \,,    \nonumber
\end{align}
where we introduced specific dimensionless combinations of the physical parameters as
\begin{align}
    L := &\ I+2\alpha\,m\,\frac{l}{a} + \alpha^2(e^2+g^2)\frac{1}{K}\,\frac{l^2}{a^2}\, , \label{defL}\\
    M := &\ \alpha\,m\,I+\alpha^2(2a^2-2l^2+e^2+g^2)\frac{l}{a}\,. \label{defM}
\end{align}
Notice that in \eqref{defL} we can alternatively employ the identity
\begin{align}
\frac{1}{K}\,\frac{l^2}{a^2} = \frac{I-\sqrt{{I^2 \mp J^2}}}{2\alpha^2(a^2-l^2)}\,.
\end{align}

A systematic, step-by-step derivation of this transformation from the Astorino form to the Pleba\'nski-Demia\'nski form of the metric is contained in Appendix~\ref{systematic-derivation}.\\

To complete the discussion of the mutual relation between the original Pleba\'nski-Demia\'nski form \eqref{PD-GP-form} of the metric PD$_{\alpha}$ (with ${\Lambda=0}$) and the new Astorino representation of this entire family of black holes, we will now substitute the parameters $k', n', \epsilon', m', e', g'$ and $\alpha'$, as given by \eqref{direct_transformation_A-PD_parameters-simplified} and \eqref{direct_transformation_A-PD-acceleraation}, into the metric functions \eqref{P'Q'eqns}, resulting in
\begin{align}
 P'=\dfrac{1}{I^2 \mp J^2}&
    \Bigg[ \dfrac{I(K-1)}{\alpha^2(a^2-l^2)}+2\,[1-\alpha^2(a^2-l^2)]\dfrac{l}{a}\,x'-I\,[K-\alpha^2(a^2-l^2)]\,x'^{\,2}\Bigg]\nonumber\\
    \times& \Bigg[ L -2M\,x' + \alpha^2\,\big[(a^2-l^2)L+(e^2+g^2)\sqrt{I^2 \mp J^2}\,\big]\,x'^{\,2}\Bigg], \label{P'-factorized}\\
 Q'=\dfrac{1}{I^2 \mp J^2}&
    \Bigg[I-2\alpha l\,[1-\alpha^2(a^2-l^2)]\,r'-\alpha^2(a^2-l^2)I\,r'^{\,2}\Bigg]\nonumber\\
    \times& \Bigg[ \dfrac{K-1}{\alpha^2(a^2-l^2)^2}\,\big[(a^2-l^2)L + (e^2+g^2)\sqrt{I^2 \mp J^2}\,\big] \nonumber
     \\& -\dfrac{2}{\alpha a}M\,r'+L\,[K-\alpha^2(a^2-l^2)]\,r'^{\,2}\Bigg]. \label{Q'-factorized}
\end{align}

We have thus arrived at a very nice result: using the new parameters introduced in \cite{Astorino:2024b, Astorino:2024a}, the quartic PD polynomials \cite{PlebanskiDemianski:1976} \emph{are factorized into the product of quadratic expressions}. Such a  factorized form is crucial for the geometrical and physical interpretation because the roots of $Q'$ and $P'$, which can now by easily found, represent the horizons and poles (axes of symmetry), respectively, of the the black-hole spacetimes. This was previously achieved in the GP and PV forms of the metric \cite{GriffithsPodolsky:2005, GriffithsPodolsky:2006, PodolskyGriffiths:2006, PodolskyVratny:2021, PodolskyVratny:2023}, but the new Astorino parametrization has now enabled us to factorize the metric functions $P'$ and $Q'$ directly in the \emph{original} PD metric.\\

There is a simplification in special cases when some of the physical parameters vanish:\\

$\bullet$ {\bf Special case  ${l=0}$: no NUT}

For black holes without the NUT twist (when ${l=0}$),
\begin{align} \label{defI-and -defJ-l=0}
    I = 1+\alpha^2a^2 \,, \qquad
    J = 0 \,,\qquad
    K = I \,,\qquad
    L = I\,,\qquad
    M = \alpha\,m\,I\,,
\end{align}
so that ${S^{-2} = \alpha^2a^2(1+\alpha^2a^2)}$ and the Pleba\'nski-Demia\'nski parameters \eqref{direct_transformation_A-PD_parameters-simplified} simplify to
\begin{align}\label{direct_transformation_A-PD_parameters-l=0}
\alpha' &= \alpha\,a \,, \nonumber\\
k'  &= 1 \,, \nonumber\\
n'  &= -\alpha\,m \,,\\
\epsilon'  &= 1 - \alpha^2(a^2+e^2+g^2)\,, \nonumber\\
\alpha' m' &= \alpha\,m \,,\nonumber\\
({e'}^2 + {g'}^2) &= a^{-2}\,({e}^2 + {g}^2)\,. \nonumber
\end{align}
The two key metric functions \eqref{P'-factorized}, \eqref{Q'-factorized} reduce to
\begin{align}
 P' &=       \big( 1 - x'^{\,2} \big)\big( 1 -2\alpha\,m\,x' + \alpha^2 (a^2+e^2+g^2)\,x'^{\,2}\big), \label{P'-factorized:l=0}\\
 Q' &=a^{-2} \big( 1 -\alpha^2a^2 r'^{\,2} \big) \big( a^2+e^2+g^2 - 2\,m\,a\,r' + a^2 r'^{\,2} \big). \label{Q'-factorized:l=0}
\end{align}
After restoring the correct physical dimensionality of the PD parameters and coordinates (as explained in Subsections~\ref{sc:dimensionality} and~\ref{sc:l=0}, in particular by applying the analogue of the relations \eqref{resc} for the choice ${\gamma=a}$) we get
${ Q' = \big( 1 -\alpha^2 r^2 \big) \big( a^2+e^2+g^2 - 2\,m\,r + r^2 \big)}$. These results fully agree with Eq.~(16.24) in Section 16.3.2 of the monograph~\cite{GriffithsPodolsky:2009}. Moreover, due to the nice factorization, the roots of $P'$ and $Q'$ --- which identify the \emph{axes} and \emph{horizons}, respectively --- can easily be determined.\\

$\bullet$ {\bf Special case ${\alpha=0}$: no acceleration}

For non-accelerating black holes, that is in the limit ${\alpha\to0}$, we get
\begin{align} \label{defI-and -defJ-alpha=0}
    I = 1\,, \qquad
    J = 0 \,,\qquad
    \frac{K-1}{\alpha^2(a^2-l^2)} = \frac{a^2-l^2}{a^2} \,,\qquad
    L = 1\,,\qquad
    M = \alpha \,m\,,
\end{align}
where the expression for $K$ can be calculated using \eqref{K-pm}, so that ${K \to 1}$, ${S^{-2} \to \alpha^2|a^2-l^2|}$, and ${\alpha' \to  \alpha\, a}$.
The PD parameters thus become
\begin{align}\label{direct_transformation_A-PD_parameters-alpha=0}
\alpha' &=\ 0 \,, \nonumber\\
k'  &= 1 - \frac{l^2}{a^2} \,, \nonumber\\
n'  &= \frac{l}{a} \,,\\
\epsilon'  &= 1\,  , \nonumber\\
m' &= \frac{m}{a}  ,\nonumber\\
{e'}^2 + {g'}^2 &= a^{-2}\,({e}^2 + {g}^2)\,. \nonumber
\end{align}
Note that setting ${l=0}$ in \eqref{direct_transformation_A-PD_parameters-alpha=0} agrees with the previous ${l=0}$ case if ${\alpha=0}$ is set in \eqref{direct_transformation_A-PD_parameters-l=0}.
The metric functions \eqref{P'-factorized}, \eqref{Q'-factorized} simplify to
\begin{align}
 P' &= 1 - \Big(\,x' - \dfrac{l}{a}\,\Big)^2, \label{P'-factorized:alpha=0}\\[1mm]
 Q' &= a^{-2} \big( a^2-l^2+e^2+g^2 - 2\,m\,a\,r' + a^2 r'^{\,2} \big). \label{Q'-factorized:alpha=0}
\end{align}
It agrees with Eq.~(16.23) in Sec.~16.3.1 of~\cite{GriffithsPodolsky:2009} after performing a shift ${x'-l/a \mapsto x'}$ in $P'$, and restoring the correct dimensionality in $Q'$ by rescaling it $a^2$ and redefining $r'$ to ${r=a\,r'}$.\\

$\bullet$ {\bf Special case ${a=0}$: no rotation}

For black holes without the Kerr-like rotation, by setting ${a = 0}$ in \eqref{I2-pm-J2}, \eqref{K-pm}, \eqref{defL} and \eqref{defM}, we get
\begin{align} \label{defI-and -defJ-a=0}
    I &=  1-\alpha^2l^2 \,, \qquad
    a^2(I^2 \mp J^2) = 4\alpha^2l^4 \,,\qquad
    aK = \alpha\,l^2   \,,  \nonumber\\
    aL&= \alpha\,\big(2ml + e^2+g^2\big)\,,\qquad
    aM = \alpha^2l\,\big(-2l^2 +e^2+g^2\big)\,.
\end{align}
Consequently, ${S^{-2} \to 2\alpha^4l^6/a^2}$ and
\begin{align}\label{direct_transformation_A-PD_parameters-a=0}
\alpha' &= \alpha^2 \,l^2 \,, \nonumber\\
k'  &= \frac{\alpha^2 l^2-1}{4\alpha^2 l^4}\,\big(2m\,l +e^2+g^2   \big) , \nonumber\\
n'  &= \frac{(m-l)\,l+(e^2+g^2)}{2 \alpha\, l^3} + \frac{\alpha}{2}(m+l) ,\\
\epsilon'  &= -2 (1+\alpha^2l^2) + \frac{1}{2}(e^2+g^2)\Big( \alpha^2 + \frac{3}{l^2}  \Big)  , \nonumber\\
\alpha' m' &= \frac{\alpha}{2}\,\Big[ -(m+l) + \alpha^2l^2(l-m)
    +\frac{e^2+g^2}{l}\,\Big] ,\nonumber\\
{e'}^2 + {g'}^2 &= \frac{1-\alpha^2l^2}{2\alpha^2 l^4}\,({e}^2 + {g}^2)\,. \nonumber
\end{align}
The two key metric functions \eqref{P'-factorized}, \eqref{Q'-factorized} thus take an explicit factorized
form
\begin{align}
 P'=\dfrac{1}{4\alpha^2l^4}&
    \bigg[ (1-\alpha^2l^2) - 2\alpha l\,(1+\alpha^2l^2)\,x' + \alpha^2l^2\,(1-\alpha^2l^2)\,x'^{\,2}\bigg]\label{P'-factorized:a=0}\\
    \times& \bigg[ - \big(2ml + e^2+g^2\big)
    + 2 \alpha l\,(-2l^2 +e^2+g^2)\,x'
    - \alpha^2l^2(-2ml +e^2+g^2)\,x'^{\,2}\bigg], \nonumber\\
 Q'=\dfrac{1}{4\alpha^2l^4}&
    \bigg[(1-\alpha^2l^2)-2\alpha l\,(1+\alpha^2l^2)\,r' + \alpha^2l^2(1-\alpha^2l^2)\,r'^{\,2}\bigg]\nonumber\\
    \times& \bigg[ (-2ml+e^2+g^2) -2 \alpha l\,(-2l^2 +e^2+g^2)\,r'
      + \alpha^2l^2(2ml + e^2+g^2)\,r'^{\,2}\bigg].
    \label{Q'-factorized:a=0}
\end{align}
This particular choice of the PD coefficients, expressed in terms of the genuine Astorino physical parameters, identifies the elusive family of accelerating NUTty black holes \emph{without the Kerr-like rotation} within the Pleba\'nski-Demia\'nski family of metrics. This will be investigated in more detail in Section~\ref{sc:a=0}. Moreover, we will present these accelerating (possibly charged) purely NUT black holes in the Griffiths-Podolsk\'y form, see the metric~\eqref{ds2_accel_NUT-rescaled}--\eqref{bar-Q-a=0}, which has not been known untill now.

\newpage

\section{Transformation to the Griffiths-Podolsk\'y metric form}
\label{sc:GP-form}

In previous section we have proven that, assuming ${\Lambda=0}$, the complete Astorino (A) class of solutions \eqref{init_metr} can be equivalently written in the Pleba\'nski-Demia\'nski (PD$_{\alpha}$) form of the metric \eqref{PD-GP-form},
\begin{align}
    \dd s^2=\dfrac{1}{{(1-\alpha'\,r'\,x')}^{\,2}}\bigg[
    &- \dfrac{Q'}{{r'}^{\,2}+{x'}^{\,2}}\,(\dd\tau'  -  {x'}^{\,2}\, \dd\phi' )^2
     + \dfrac{P'}{{r'}^{\,2}+{x'}^{\,2}}\,(\dd\tau'  +  {r'}^{\,2}\, \dd\phi' )^2 \nonumber\\
    &+ c^2({r'}^{\,2}+{x'}^{\,2}) \Big(\,\dfrac{\dd {r'}^{\,2}}{Q'} + \dfrac{\dd {x'}^{\,2}}{P'}\,\Big)\bigg], \label{PD-GP-form-again}
\end{align}
where the metric functions $P'(x')$ and $Q'(r')$, explicitly expressed in terms of the very convenient Astorino parameters and factorized, are generally given by \eqref{P'-factorized} and \eqref{Q'-factorized}, respectively.

Our aim in this section is to relate the PD$_{\alpha}$ form \eqref{PD-GP-form-again} of the type~D black holes to the Griffiths-Podolsk\'y (GP) form of these solutions, summarized in \cite{GriffithsPodolsky:2009}. This will elucidate the \emph{direct relation} of the GP metric to the A metric. In particular, it will clarify the relation between the Astorino initial parameters $\alpha,~a,~l,~m,~e,~g$ and the physical parameters $\tilde{\alpha},~\tilde{a},~\tilde{l},~\tilde{m},~\tilde{e},~\tilde{g}$ employed in the GP form of the metric previously \cite{GriffithsPodolsky:2005, GriffithsPodolsky:2006, PodolskyGriffiths:2006}.

We start by performing a simple rescaling of coordinates, bringing the PD$_{\alpha}$ metric to the PD$_{\alpha\omega}$ metric \eqref{PleDemMetric} which involves an additional \emph{twist parameter}~$\omega$, namely\footnote{Note that the acceleration parameter $\alpha'$ is already included in the metric PD$_{\alpha}$ so we do not have to include it in the coordinate transformation as was done \cite{GriffithsPodolsky:2006} and repeated at the beginning of section~\ref{sc:PD-form}.}
\begin{equation}
    x' \mapsto \sqrt{\omega}\,x',\qquad
    r' \mapsto \dfrac{r'}{\sqrt{\omega}},\qquad
 \tau' \mapsto \sqrt{\omega}\,\tau',\qquad
 \phi' \mapsto \sqrt{\omega}\,\phi'.\label{coord_tr_1}
\end{equation}
These rescalings bring the metric \eqref{PD-GP-form-again} to the PD$_{\alpha\omega}$ form
  \begin{eqnarray}\label{PleDemMetric-again}
  &&\hskip-3pc\dd s^2=\frac{1}{(1-\alpha'\, r'\,x')^2} \Bigg[
 -\frac{{\cal Q}}{r'^{\,2}+\omega^2x'^{\,2}}(\dd\tau'-\omega x'^{\,2}\dd\phi')^2
 +\frac{r'^{\,2}+\omega^2x'^{\,2}}{{\cal Q}}\,\dd r'^{\,2} \nonumber \\
  &&\hskip6pc
  +\frac{{\cal P}}{r'^{\,2}+\omega^2x'^{\,2}}(\omega\dd\tau'+r'^{\,2}\dd\phi')^2
  +\frac{r'^{\,2}+\omega^2x'^{\,2}}{{\cal P}}\,\dd x'^{\,2} \Bigg],
  \end{eqnarray}
where
${\mathcal{P}(x') := P'(\sqrt{\omega}\,x')}$ and
${\mathcal{Q}(r') := \omega^2\,Q'\Big(\dfrac{r'}{\sqrt{\omega}}\Big)}$.\\

Following \cite{GriffithsPodolsky:2005, GriffithsPodolsky:2006, GriffithsPodolsky:2009}, the next step is to perform a coordinate transformation
\begin{equation}
    x'=\dfrac{\tilde{l}}{\omega}+\dfrac{\tilde{a}}{\omega}\,\tilde{x},\qquad
 \tau'=t-\dfrac{(\tilde{a}+\tilde{l})^2}{\tilde{a}}\,\varphi,\qquad
 \phi'=-\dfrac{\omega}{\tilde{a}}\,\varphi,\qquad
    r'=\tilde{r},\label{coord_tr_2}
\end{equation}
where $\tilde{a}$ represents the Kerr-like rotational parameter, while $\tilde{l}$ represent the NUT-like parameter. After these linear transformations, the metric becomes
\begin{align}
    \dd s^2=\dfrac{1}{{\tilde{\Omega}}^2}
    \Bigg[&-\dfrac{\tilde{\mathcal{Q}}}{\,\tilde{\rho}^{\,2}}
    \Big[\dd t-\big(\tilde{a}(1-\tilde{x}^2)+2\tilde{l}(1-\tilde{x})\big)\dd\varphi\Big]^2
     +\dfrac{\,\tilde{\rho}^{\,2}}{\tilde{\mathcal{Q}}}\,\dd\tilde{r}^2\nonumber\\
    &+\dfrac{\,\tilde{\rho}^{\,2}}{\tilde{\mathcal{P}}}\,\dd\tilde{x}^2
     +\dfrac{\tilde{\mathcal{P}}}{\,\tilde{\rho}^{\,2}}
    \big[\tilde{a}\,\dd t-\big(\tilde{r}^2+(\tilde{a}+\tilde{l})^2\big)\,\dd\varphi\big]^2\Bigg],
    \label{ds2_accel_kerr_new}
\end{align}
in which
\begin{eqnarray}
\tilde{\Omega} :=& 1-\dfrac{\alpha'}{\omega}\,\tilde{r}\,(\tilde{l}+\tilde{a}\,\tilde{x}), \\[2mm]
\tilde{\rho}^{\,2} :=& \tilde{r}^{\,2}+(\tilde{l}+\tilde{a}\,\tilde{x})^2, \\[2mm]
\tilde{\mathcal{P}}(\tilde{x}) :=& \dfrac{\omega^2}{\tilde{a}^2}\,P'\Big(\dfrac{\tilde{l}+\tilde{a}\,\tilde{x}}{\sqrt{\omega}}\Big),
\label{mathcalP}\\[1mm]
\tilde{\mathcal{Q}}(\tilde{r}) :=& \hspace{-4mm} \omega^2\, Q' \Big(\dfrac{\tilde{r}}{\sqrt{\omega}}\Big)\,,
\label{mathcalQ}
\end{eqnarray}
where the functions $P'(x'), Q'(r')$ are given by \eqref{P'-factorized}, \eqref{Q'-factorized}. This is the general Griffiths-Podolsk\'y form of the metric, as summarized in Eq.~(16.12) in~\cite{GriffithsPodolsky:2009}.\\

By inspecting the conformal factor $\tilde{\Omega}$ we observe that the \emph{GP acceleration parameter $\tilde{\alpha}$ is equal to the PD acceleration parameter}, ${\tilde{\alpha}=\alpha'}$. Using \eqref{direct_transformation_A-PD-acceleraation}, we can thus directly relate the GP acceleration to the  A parameters as
\begin{equation}
    \tilde{\alpha} = \alpha\,a\,\big[\,K - \alpha^2(a^2-l^2)\big].
    \label{tild_alph}
\end{equation}
Expressed explicitly \emph{in terms of the new Astorino parameters}, this is actually quite an involved expression \eqref{direct_transformation_A-PD-acceleration-explicit}, namely
\begin{equation}\label{direct_transformation_A-PD-acceleration-explicit-again}
 \tilde{\alpha} = \alpha\,\,\tfrac{1}{2}\big[\,a - \alpha^2 a (a^2-l^2)
+ \sqrt{a^2+\alpha^4 a^2(a^2-l^2)^2+2\alpha^2(a^2-l^2)(a^2-2l^2)}\, \big].
\end{equation}
Clearly, ${\alpha=0}$ implies ${\tilde{\alpha}=0}$, which is expected. However, by setting ${l=0}$ we get ${\tilde{\alpha}=\alpha\,a}$. It means that the GP acceleration parameter $\tilde{\alpha}$ \emph{also vanishes for} ${l=0=a}$. Similarly, for ${a=0}$ we get ${ \tilde{\alpha}=\alpha^2 l^2}$, so that
$\tilde{\alpha}$ \emph{also vanishes for} ${a=0=l}$. This degeneracy is an unfortunate feature of the original GP representation of the whole class of type~D black holes, preventing to identify the genuine subclass of accelerating NUT black holes without the Kerr-like rotation --- which exists, and is nicely contained in the Astorino metric \eqref{init_metr}--\eqref{delta_x_init}.\\

Let us now concentrate on the GP ``rotational'' parameters $\tilde{a}$ and $\tilde{l}$. In the transformation \eqref{coord_tr_2} these are arbitrary constants. However, they are naturally constrained by the requirement that in the final form of the GP metric the spherical-like coordinate $\theta$ should be introduced instead of $\tilde{x}$ in \eqref{ds2_accel_kerr_new} via the relation ${\tilde{x}=\cos\theta}$. To this end, the function $\tilde{\mathcal{P}}(\tilde{x})$ must be written in the specific factorized form
\begin{equation}
\tilde{\mathcal{P}}=(1-\tilde{x}^2)(1-a_3\,\tilde{x}-a_4\,\tilde{x}^2),\label{tild_P_simpl}
\end{equation}
and this can be achieved by a unique values of the parameters $\tilde{a}$ and $\tilde{l}$. Indeed, for the choice\footnote{In view of the definition \eqref{defI-and -defJ}, the upper sign applies for ${a^2>l^2}$, while the lower sign applies for ${a^2<l^2}$.}
\begin{align}
    \tilde{a}&= \dfrac{ \sqrt{\omega}}{K-\alpha^2(a^2-l^2)}\,
              \frac{\sqrt{I^2 \mp J^2}}{I}\,,\label{a_tld}\\
    \tilde{l}&= \dfrac{ \sqrt{\omega}}{K-\alpha^2(a^2-l^2)}\,
              F\,\dfrac{l}{a}\,, \label{l_tld}
\end{align}
where $F$ denotes the fraction
\begin{equation}\label{def-F}
F := \dfrac{1-\alpha^2(a^2-l^2)}{1+\alpha^2(a^2-l^2)},
\end{equation}
the first bracket in $P'$ given by \eqref{P'-factorized}, expressed in the new coordinate $\tilde{x}$ such that ${x'=(\tilde{l}+\tilde{a}\,\tilde{x})/\sqrt{\omega}}$, becomes
\begin{align}
\dfrac{I(K-1)}{\alpha^2(a^2-l^2)}+2\,[1-\alpha^2(a^2-l^2)]&\dfrac{l}{a}\,x'-I\,[K-\alpha^2(a^2-l^2)]\,x'^{\,2}\nonumber\\
 & = \dfrac{I^2 \mp J^2}{I\,[K-\alpha^2(a^2-l^2)]}\,(1-\tilde{x}^2)\,.
\end{align}
Notice also that this natural fixing of $\tilde{a}$ and $\tilde{l}$ can be rewritten using the relation \eqref{tild_alph} as
\begin{align}
    \tilde{\alpha}\,\tilde{a}&= \alpha\,a\,\sqrt{\omega}\, \dfrac{\sqrt{I^2 \mp J^2}}{I}\,, \label{a_tld-alter}\\
    \tilde{\alpha}\,\tilde{l}&= \alpha\,l\,\sqrt{\omega}\, F\,. \label{l_tld-alter}
\end{align}

Evaluating also the second bracket in $P'$, we obtain the metric function $\tilde{\mathcal{P}}$ in   \eqref{ds2_accel_kerr_new}
\begin{align}
 \tilde{\mathcal{P}}(\tilde{x})=&\ (1-\tilde{x}^2)\,\dfrac{\omega^2}{\tilde{a}^2}\,\dfrac{1}{I\,[K-\alpha^2(a^2-l^2)]}\,\nonumber\\
    & \times \Bigg[\,L  - 2M\dfrac{\tilde{l}}{\sqrt{\omega}}
       +\alpha^2\big[(a^2-l^2)L+(e^2+g^2)\,\sqrt{I^2 \mp J^2}\,\big] \dfrac{\tilde{l}^2}{\omega}\nonumber\\
    &\qquad
       -\Big(2M\dfrac{\tilde{a}}{\sqrt{\omega}}-2\alpha^2\big[(a^2-l^2)L+(e^2+g^2)\,\sqrt{I^2 \mp J^2}\,\big]
       \dfrac{\tilde{l}\tilde{a}}{\omega}\Big)\,\tilde{x}\nonumber\\
    &\qquad
       +\alpha^2\,\dfrac{\tilde{a}^2}{\omega}\big[(a^2-l^2)L+(e^2+g^2)\,\sqrt{I^2 \mp J^2}\,\big]\,\tilde{x}^2 \Bigg],\label{tild_P_x}
\end{align}
which is indeed of the required factorized form \eqref{tild_P_simpl} ---  up to an overall rescaling which can always be achieved.

Indeed, so far $\omega$ has been a \emph{free ``twist'' parameter} introduced by \eqref{coord_tr_1}. To describe a black hole with the horizon topology of a sphere, the function $\tilde{\mathcal{P}}(\tilde{x})$ has to satisfy the condition $\tilde{\mathcal{P}}(\tilde{x}=0)=1$ which directly follows from \eqref{tild_P_simpl} (see also \cite{GriffithsPodolsky:2006}). For \eqref{tild_P_x}, using \eqref{a_tld-alter}, \eqref{l_tld-alter} and \eqref{tild_alph}, we thus derive a \emph{special value of} $\omega$, namely\footnote{Let us note, however, that in  Sec.~\ref{sc:issue-twist} we will demonstrate that an arbitrary value of $\omega$ in the metric can be restored by a rescaling of the acceleration parameter.}
\begin{align}
    \omega_0 =
    \dfrac{\alpha\,a}{\tilde{\alpha}}\,\dfrac{I^2 \mp J^2}{I}
    \Big( L - 2M\dfrac{\alpha\,l}{\tilde{\alpha}}\,F
    +\dfrac{\alpha^4l^2}{\tilde{\alpha}^2}\Big[(a^2-l^2)L+(e^2+g^2)\,\sqrt{I^2 \mp J^2}\Big] F^2
    \Big)^{-1}.\label{omega_0}
\end{align}
This brings $\tilde{\mathcal{P}}$ exactly to the desired form (\ref{tild_P_simpl}), that is
\begin{equation}
\tilde{\mathcal{P}}(\tilde{x})=(1-\tilde{x}^2)\,\tilde{P}(\tilde{x}),
\qquad\hbox{where}\qquad
\tilde{P}(\tilde{x}) :=  1-a_3\,\tilde{x}-a_4\,\tilde{x}^2,
\label{tild_P_simpl-again}
\end{equation}
with the coefficients $a_3$ and $a_4$ given by
\begin{align}
    a_3=&\dfrac{2}{\sqrt{I^2 \mp J^2}}\,\omega_0
    \Big( M - \dfrac{\alpha^3 l }{\tilde{\alpha}}\Big[(a^2-l^2)L+(e^2+g^2)\,\sqrt{I^2 \mp J^2}\,\Big] F   \Big),\\
    a_4=&-\dfrac{\alpha^3 a}{\tilde{\alpha}\, I}\,\omega_0\Big[(a^2-l^2)L+(e^2+g^2)\,\sqrt{I^2 \mp J^2}\,\Big].
\end{align}

The last metric function $\tilde{\mathcal{Q}}(\tilde{r})$ easily follows from the relations \eqref{mathcalQ} and \eqref{P'Q'eqns},
\begin{align}
    \tilde{\mathcal{Q}}(\tilde{r})=
    \omega_0^2(k'+{e'}^{\,2}+{g'}^{\,2})-2\,\omega_0^{3/2}m'\,\tilde{r}+\omega_0\epsilon'\,\tilde{r}^2
    -2\sqrt{\omega_0}\,\alpha'n'\,\tilde{r}^3-{\alpha'}^{\,2}k'\,\tilde{r}^4\,,
\end{align}
recalling that ${\tilde{\alpha}=\alpha'}$, and introducing the rescaled parameters
\begin{align}\label{rescaled-GP}
    \tilde{k}:=k',\quad
    \tilde{m}:=\omega_0^{3/2}\,m',\quad
    \tilde{n}:=\omega_0^{3/2}\,n'\,,\quad
    \tilde{\epsilon}:=\omega_0\,\epsilon',\quad
    \tilde{e}:=\omega_0\,e',\quad
    \tilde{g}:=\omega_0\,g',
\end{align}
where ${k', n', \epsilon', m', e', g'}$ are given by \eqref{direct_transformation_A-PD_parameters-simplified}.
\newpage

To obtain the Griffiths-Podolsk\'y form of the metric it now suffices to introduce the angular coordinate
$\theta$ in \eqref{ds2_accel_kerr_new} via the simple relation ${\tilde{x}=\cos\theta}$, resulting in
\begin{align}
    \dd s^2=\dfrac{1}{{\tilde{\Omega}}^2}
    \Bigg[&-\dfrac{\tilde{Q}}{\,\tilde{\rho}^{\,2}}
    \Big[\dd t-\big(\tilde{a}\sin^2\theta+2\tilde{l}(1-\cos\theta)\big)\dd\varphi\Big]^2
     +\dfrac{\,\tilde{\rho}^{\,2}}{\tilde{Q}}\,\dd\tilde{r}^2\nonumber\\
    &+\dfrac{\,\tilde{\rho}^{\,2}}{\tilde{P}}\,\dd\theta^2
     +\dfrac{\tilde{P}}{\,\tilde{\rho}^{\,2}}\sin^2\theta\,
    \big[\,\tilde{a}\,\dd t-\big(\tilde{r}^2+(\tilde{a}+\tilde{l})^2\big)\,\dd\varphi\big]^2\Bigg],
    \label{ds2_accel_kerr_new_polar}
\end{align}
where
\begin{align}
  \tilde{\Omega}       &= 1-\dfrac{\tilde{\alpha}}{\omega_0}\,\tilde{r}\,(\tilde{l}+\tilde{a} \cos\theta), \label{tilde-Omega-x}\\
  \tilde{\rho}^{\,2}   &= \tilde{r}^{\,2}+(\tilde{l}+\tilde{a}\,\cos\theta)^2, \label{tilde-rho}\\
  \tilde{P}(\theta)    &= 1-a_3\cos\theta-a_4\cos^2\theta\,, \label{tilde-P}\\
  \tilde{Q}(\tilde{r}) &= (\omega_0^2\,\tilde{k}+\tilde{e}^{\,2}+\tilde{g}^{\,2})
        -2\tilde{m}\,\tilde{r}+\tilde{\epsilon}\,\tilde{r}^{\,2}
        -2\tilde{\alpha}\dfrac{\tilde{n}}{\omega_0}\,\tilde{r}^{\,3}
        -\tilde{\alpha}^2\tilde{k}\,\tilde{r}^{\,4}\,. \label{tilde-Q}
\end{align}
It is exactly the GP metric given by Eqs.~(6.18), (6.19) in the monograph~\cite{GriffithsPodolsky:2009}.

Moreover, a straightforward (but somewhat lengthy) calculation proves that the above parameters satisfy the following set of relations
\begin{align}
  a_3 &= 2\tilde{\alpha}\dfrac{\tilde{a}}{\omega_0}\tilde{m}
      -4\tilde{\alpha}^2\dfrac{\tilde{a}\tilde{l}}{\omega_0^2}(\omega_0^2\tilde{k}+\tilde{e}^{\,2}+\tilde{g}^{\,2}),
      \label{GP_a3_rel}\\
  a_4 &=
      -\tilde{\alpha}^2\dfrac{\tilde{a}^2}{\omega_0^2}(\omega_0^2\tilde{k}+\tilde{e}^{\,2}+\tilde{g}^{\,2}),
      \label{GP_a4_rel}\\[1mm]
  \tilde{\epsilon} &= \dfrac{\omega_0^2 \tilde{k}}{\tilde{a}^2
      -\tilde{l}^2}+4\tilde{\alpha}\dfrac{\tilde{l}}{\omega_0}\tilde{m}
      -(\tilde{a}^2+3\tilde{l}^2)\Big[\,\dfrac{\tilde{\alpha}^2}{\omega_0^2}(\omega_0^2\tilde{k}+\tilde{e}^{\,2}+\tilde{g}^{\,2})\Big],
      \label{GP_eps_rel}\\[1mm]
  \tilde{n} &= \dfrac{\omega_0^2 \tilde{k}\tilde{l}}{\tilde{a}^2-\tilde{l}^2}
      -\tilde{\alpha}\dfrac{\tilde{a}^2-\tilde{l}^2}{\omega_0}\tilde{m}
      +(\tilde{a}^2-\tilde{l}^2)\tilde{l}\Big[\,\dfrac{\tilde{\alpha}^2}{\omega_0^2}(\omega_0^2\tilde{k}+\tilde{e}^{\,2}+\tilde{g}^{\,2})\Big],
      \label{GP_n_rel}\\[1mm]
  \Big(\dfrac{\omega_0^2}{\tilde{a}^2-\tilde{l}^2}+3\tilde{\alpha}^2\,\tilde{l}^{\,2}\Big)\tilde{k} &=
      1+2\tilde{\alpha}\dfrac{\tilde{l}}{\omega_0}\tilde{m}
      -3\tilde{\alpha}^2\dfrac{\tilde{l}^{\,2}}{\omega_0^2}(\tilde{e}^{\,2}+\tilde{g}^{\,2}),
      \label{GP_k_rel}
\end{align}
which are exactly the expressions in Eqs.~(16.20) and~(16.15)--(16.17) in~\cite{GriffithsPodolsky:2009}. \\

This finishes the construction of the Griffiths-Podolsk\'y form of the general metric of black holes of algebraic type~D. Moreover, it explicitly \emph{demonstrates the full equivalence of the GP form with the PD and A forms} of this large class of spacetimes.

However, it should be emphasized that there is a subtle but very important difference: the original Griffiths-Podolsk\'y physical parameters ${\tilde{\alpha}, \tilde{a}, \tilde{l}}$ (representing acceleration, Kerr-like rotation, NUT twist) and ${\tilde{m}, \tilde{e}, \tilde{g}}$ (representing mass, electric charge, magnetic charge) in the metric \eqref{ds2_accel_kerr_new_polar} \emph{are not} the new Astorino \emph{genuine parameters} ${\alpha, a, l}$ and ${m, e, g}$, which properly separate the corresponding subclasses when they are set to zero.

In fact, the old GP~parameters are now explicitly expressed in terms of the new A~parameters \emph{via the complicated relations} \eqref{direct_transformation_A-PD-acceleration-explicit-again}, \eqref{a_tld}, \eqref{l_tld} and \eqref{rescaled-GP} with \eqref{direct_transformation_A-PD_parameters-simplified}. Moreover, the additional twist parameter $\omega$ has a very special value $\omega_0$ given by \eqref{omega_0}. Due to their highly involved structure, these could not be guessed in previous investigation of this family of spacetimes.

To elucidate the relation between the GP and A (that is also A$^+$) physical parameters in more detail, it seems to be instructive to \emph{consider the special cases} ${\alpha=0}$, ${l=0}$, ${a=0}$. It will clearly demonstrate in which situations the two sets of parameters agree, and what are their specific differences.

\section{Special cases}
\label{sc:the-special-cases}

To analyze various special subcases of black holes, it is first necessary to clarify the \emph{freedom in the choice of the twist parameter}~$\omega$, and also to consider the \emph{physical dimensionality of the parameters}~$\tilde{\alpha}$, $\tilde{l}$ and~$\tilde{a}$.

\subsection{An issue of the twist parameter $\omega$}
\label{sc:issue-twist}

In previous section, an explicit transformation from the Astorino form of the metric to the Griffiths-Podolsk\'y one was presented. It involves a very special, \emph{unique choice} of the twist parameter ${\omega=\omega_0}$ given by \eqref{omega_0}. This may be seen as a contradiction to statements in the published  works, such as \cite{GriffithsPodolsky:2006}, where it was argued that $\omega$ is a \emph{free parameter} (with a general restriction that it is related to the twist of the congruence generated by PNDs, and thus to both the Kerr-like rotational parameter~$\tilde{a}$ and the NUT-like parameter $\tilde{l}$.

However, it can be demonstrated that there is no such contradiction because arbitrary (nonzero) value of $\omega$ can be restored from (nonzero) $\omega_0$. This is achieved by a simple \emph{rescaling of the acceleration parameter},
\begin{equation}\label{rescaling-of-alpha}
    \tilde{\alpha} \quad\mapsto\quad \tilde{\alpha}\,\dfrac{\omega_0}{\omega}\,,
\end{equation}
while keeping all other physical parameters (that is $\tilde{a},~\tilde{l},~\tilde{m},~\tilde{e},~\tilde{g}$) the same. After this substitution, the set of relations \eqref{GP_a3_rel}--\eqref{GP_k_rel} become
\begin{align}
  a_3 &= 2\tilde{\alpha}\dfrac{\tilde{a}}{\omega}\tilde{m}
      -4\tilde{\alpha}^2\dfrac{\tilde{a}\tilde{l}}{\omega^2}(\omega_0^2\tilde{k}+\tilde{e}^{\,2}+\tilde{g}^{\,2}),
      \label{GP_a3_rel-omega}\\
  a_4 &=
      -\tilde{\alpha}^2\dfrac{\tilde{a}^2}{\omega^2}(\omega_0^2\tilde{k}+\tilde{e}^{\,2}+\tilde{g}^{\,2}),
      \label{GP_a4_rel-omega}\\[1mm]
  \tilde{\epsilon} &= \dfrac{\omega_0^2 \tilde{k}}{\tilde{a}^2
      -\tilde{l}^2}+4\tilde{\alpha}\dfrac{\tilde{l}}{\omega}\tilde{m}
      -(\tilde{a}^2+3\tilde{l}^2)\Big[\,\dfrac{\tilde{\alpha}^2}{\omega^2}(\omega_0^2\tilde{k}+\tilde{e}^{\,2}+\tilde{g}^{\,2})\Big],
      \label{GP_eps_rel-omega}\\[1mm]
  \tilde{n} &= \dfrac{\omega_0^2 \tilde{k}\tilde{l}}{\tilde{a}^2-\tilde{l}^2}
      -\tilde{\alpha}\dfrac{\tilde{a}^2-\tilde{l}^2}{\omega}\tilde{m}
      +(\tilde{a}^2-\tilde{l}^2)\tilde{l}\Big[\,\dfrac{\tilde{\alpha}^2}{\omega^2}(\omega_0^2\tilde{k}+\tilde{e}^{\,2}+\tilde{g}^{\,2})\Big],
      \label{GP_n_rel-omega}\\[1mm]
  \Big(\dfrac{\omega^2}{\tilde{a}^2-\tilde{l}^2}+3\tilde{\alpha}^2\,\tilde{l}^{\,2}\Big)\,\frac{\omega_0^2}{\omega^2}\,\tilde{k} &=
      1+2\tilde{\alpha}\dfrac{\tilde{l}}{\omega}\tilde{m}
      -3\tilde{\alpha}^2\dfrac{\tilde{l}^{\,2}}{\omega^2}(\tilde{e}^{\,2}+\tilde{g}^{\,2}).
      \label{GP_k_rel-omega}
\end{align}
The last equation suggest a rescaling
\begin{equation}\label{rescaling-of-k}
    \tilde{k} \quad\mapsto\quad \dfrac{\omega^2}{\omega_0^2}\,\tilde{k}\,,
\end{equation}
which replaces the special value of $\omega_0$ in all relations \eqref{GP_a3_rel-omega}--\eqref{GP_k_rel-omega} by an arbitrary value~$\omega$ (because $\omega_0^2\tilde{k}$ is replaced by $\omega^2\tilde{k}$).

Concerning the metric functions given by \eqref{tilde-Omega-x}--\eqref{tilde-Q}, $\tilde{\rho}^{\,2}$ and $\tilde{P}$ remain the same, while $\tilde{\Omega} $ and $\tilde{Q}$ change to
\begin{align}
  \tilde{\Omega}       &= 1-\dfrac{\tilde{\alpha}}{\omega}\,\tilde{r}\,(\tilde{l}+\tilde{a} \cos\theta), \label{tilde-Omega-omega}\\
  \tilde{Q}(\tilde{r}) &= (\omega^2\tilde{k}+\tilde{e}^{\,2}+\tilde{g}^{\,2})
        -2\tilde{m}\,\tilde{r}+\tilde{\epsilon}\,\tilde{r}^{\,2}
        -2\tilde{\alpha}\dfrac{\tilde{n}}{\omega}\,\tilde{r}^{\,3}
        -\tilde{\alpha}^2\tilde{k}\,\tilde{r}^{\,4}\,. \label{tilde-Q-omega}
\end{align}
The metric \eqref{ds2_accel_kerr_new_polar} thus takes exactly the form of Eq.~(16.18)--(16.20) in~\cite{GriffithsPodolsky:2009}, that is the original Griffiths-Podolsk\'y metric with a \emph{general} (not fixed) value of $\omega$.

To conclude, the simple rescaling \eqref{rescaling-of-alpha} of the acceleration parameter $\tilde{\alpha}$, accompanied by the rescaling \eqref{rescaling-of-k} of the parameter $\tilde{k}$, \emph{restores arbitrariness of} $\omega$ in the GP metric. In other words, the original GP metrics with \emph{different values of~$\omega$ are equivalent} (unless ${\omega=0}$).

\subsection{Restoring the physical dimensionality of the black-hole parameters}
\label{sc:dimensionality}

Recall that it is convenient and natural to consider that \emph{all the coordinates and genuine physical parameters in the Astorino metric} \eqref{init_metr}--\eqref{delta_x_init} \emph{have the usual physical dimension} --- as in the Boyer-Lindquist-type coordinates for the Kerr-Newman black holes (and thus their generalization in the Griffiths-Podolsk\'y metric form). In particular, the physical parameters $m, a, l, e, g$ have the dimension of length (while $\alpha$ has the dimension 1/length). Also the coordinates $r$ and $t$ have the dimension of length.

On the other hand, the GP parameters ${\tilde{m}, \tilde{\alpha}, \tilde{a}, \tilde{l}, \tilde{e}, \tilde{g}}$ in the metric \eqref{ds2_accel_kerr_new_polar} \emph{are dimensionless}. The reason is that they have been obtained  from the dimensionless PD coefficients ${k', n', \epsilon', m', e', g'}$ --- given by \eqref{direct_transformation_A-PD_parameters-simplified} --- using the relations \eqref{rescaled-GP} and  ${\tilde{\alpha}=\alpha'}$.

Their proper physical dimensionality can be restored by introducing a \emph{parameter~$\gamma$ with the dimension of length}, namely by rescaling the GP parameters in such a way that
\begin{align}
    & \tilde{r} \mapsto \gamma\,\tilde{r}, \qquad
      \tilde{m} \mapsto \gamma\,\tilde{m}, \qquad
      \tilde{a} \mapsto \gamma\,\tilde{a}, \qquad
      \tilde{l} \mapsto \gamma\,\tilde{l}, \qquad
      \omega \mapsto \gamma\,\omega, \nonumber\\
    &
      \tilde{e} \mapsto \gamma\,\tilde{e}, \qquad
      \tilde{g} \mapsto \gamma\,\tilde{g}, \qquad
      \tilde{\alpha} \mapsto \dfrac{\tilde{\alpha}}{\gamma}, \qquad
    \tilde{Q} \mapsto \gamma^2 \,\tilde{Q}\,. \label{resc}
\end{align}
Let us note that after this rescaling the conformal factor $\tilde{\Omega}$ given by  \eqref{tilde-Omega-omega} remains the same, while \eqref{tilde-Omega-x} changes to
\begin{align} \label{tilde-Omega-omega-gamma}
\tilde{\Omega} = 1-\frac{\tilde{\alpha}}{\gamma\omega_0}\,\tilde{r}\,(\tilde{l}+\tilde{a} \cos\theta)
\end{align}
if we keep $\omega_0$, fixed by \eqref{omega_0}, dimensionless.

In the most general case, it is not a~priori clear how to choose the unique value of $\gamma$. However, it can be easily identified in the particular cases of black holes, to recover the standard forms of these well-known solution.

\subsection {The special case ${l=0}$: no NUT}
\label{sc:l=0}

Using \eqref{defI-and -defJ-l=0}, from \eqref{omega_0} we get ${\omega_0 =
    \alpha\,a / \tilde{\alpha}}$. Recalling \eqref{direct_transformation_A-PD-acceleration-explicit-again}, which gives the relation
\begin{align}
\tilde{\alpha} &=\ \alpha\,a,
\end{align}
we immediately obtain a nice result ${\omega_0 = 1 }$. The expressions \eqref{a_tld}, \eqref{l_tld} then reduce to
\begin{align}
    \tilde{a} = \sqrt{\omega_0} = 1\,, \qquad
    \tilde{l} = 0 , \label{l_tld-l=0}
\end{align}
and the coefficients $a_3$ and $a_4$ are simply ${a_3 = 2\alpha m}$, ${a_4 = -\alpha^2 (a^2+e^2+g^2)}$,
so that
\begin{align}
  \tilde{P} = 1 - 2\alpha m \,\cos\theta + \alpha^2 (a^2+e^2+g^2) \cos^2\theta   \,.
  \label{tilde-P-Al=0}
\end{align}
The function $\tilde{Q}$ is given by \eqref{tilde-Q} with the coefficients determined by \eqref{rescaled-GP} and \eqref{direct_transformation_A-PD_parameters-l=0},
\begin{align}
 \tilde{Q} = \frac{1}{a^2} \big((a^2+e^2+g^2) - 2m\,a\,\tilde{r} +a^2\tilde{r}^{\,2} \big)
                           \big(1 - \alpha^2a^2\,\tilde{r}^{\,2}\big). \label{tilde-Q-Al=0}
\end{align}
Together with
\begin{align}
  \tilde{\Omega}      = 1-\alpha\,a\,\tilde{r} \cos\theta, \qquad
  \tilde{\rho}^{\,2}  = \tilde{r}^{\,2}+\cos^2\theta, \label{tilde-Omega-rho-Al=0}
\end{align}
it gives an explicit form of the metric \eqref{ds2_accel_kerr_new_polar} in terms of the \emph{Astorino physical parameters}.

Restoring now the \emph{correct dimensionality of the PD parameters} by using \eqref{resc} with the simple choice
\begin{equation}\label{gamma_for_l=0}
    \gamma = a,
\end{equation}
we obtain
\begin{align}
    \tilde{a}=a, \qquad
    \tilde{l}=0, \qquad
    \tilde{\alpha}=\alpha
\end{align}
The metric functions take the form
\begin{align}
  \tilde{\Omega}       &= 1-\tilde{\alpha}\,\tilde{r} \cos\theta, \nonumber\\
  \tilde{\rho}^{\,2}   &= \tilde{r}^{\,2}+\tilde{a}^2 \cos^2\theta, \nonumber\\
  \tilde{P}            &= 1 - 2\tilde{\alpha} \tilde{m} \,\cos\theta + \tilde{\alpha}^{\,2} (\tilde{a}^{\,2}+\tilde{e}^{\,2}+\tilde{g}^{\,2}) \cos^2\theta ,\nonumber\\
  \tilde{Q} &=  \big((\tilde{a}^{\,2}+\tilde{e}^{\,2}+\tilde{g}^{\,2}) - 2\tilde{m}\,\tilde{r} + \tilde{r}^{\,2} \big) \big(1 - \tilde{\alpha}^{\,2}\tilde{r}^{\,2}\big), \label{tilde-Q-l=0}
\end{align}
which fully agrees with previous GP and PV forms of this class of accelerating Kerr-Newman black holes without the NUT parameter, as presented in Eq.~(35)--(39) of \cite{PodolskyVratny:2021} and also in Sec.~16.3.2 of~\cite{GriffithsPodolsky:2009}. In this standard form of the metric, any remaining physical parameters can be set to zero, in any order.

For ${\tilde{m}^{\,2}>\tilde{a}^{\,2}+\tilde{e}^{\,2}+\tilde{g}^{\,2}}$, both the key metric functions can be written in a factorized form
\begin{align}
  \tilde{P} &= \big(1 - \tilde{\alpha}\, \tilde{r}_+\,\cos\theta\big)\big(1 - \tilde{\alpha}\, \tilde{r}_-\,\cos\theta\big)  \,,\\
  \tilde{Q} &= \big( \tilde{r} - \tilde{r}_+ \big)\big( \tilde{r} - \tilde{r}_- \big)
                             \big(1 - \tilde{\alpha}\,\tilde{r}\big)\big(1 + \tilde{\alpha}\,\tilde{r} \big) \,,
\end{align}
where ${\tilde{r}_\pm = \tilde{m} \pm \sqrt{\tilde{m}^{\,2}-\tilde{a}^{\,2}-\tilde{e}^{\,2}-\tilde{g}^{\,2}}}$. The roots of $\tilde{Q}$ define the position of the horizons. As explained in detail in~\cite{PodolskyVratny:2021}, there are two black-hole horizons and two acceleration horizons. Extremal black holes with a degenerate horizon occur when ${\tilde{m}^{\,2}=\tilde{a}^{\,2}+\tilde{e}^{\,2}+\tilde{g}^{\,2}}$, while for ${\tilde{\alpha}=0}$ the acceleration horizons disappear.

\newpage

\subsection {The special case ${\alpha=0}$: no acceleration}
\label{sc:alpha=0}

In this case
\begin{align}
    I = 1\,, \qquad
    J = 0 \,,\qquad
    K = 1 \,,\qquad
    L = 1\,,\qquad
    M = \alpha \,m\,,
\end{align}
see \eqref{defI-and -defJ-alpha=0}. From \eqref{direct_transformation_A-PD-acceleration-explicit-again} we get the limit ${\tilde{\alpha} \to \alpha\,a}$, so that using \eqref{omega_0} we obtain
\begin{align}
    \tilde{\alpha}=0\,,\qquad
    \omega_0 = 1 .\label{omega_0-for-alpha=0}
\end{align}
Consequently, the relations \eqref{a_tld}, \eqref{l_tld} give
\begin{align}
    \tilde{a} = 1\,, \qquad
    \tilde{l} = \frac{l}{a} , \label{l_tld-alpha=0}
\end{align}
and the coefficients $a_3$ and $a_4$ are simply ${a_3 = 0 = a_4}$. Therefore, employing
\eqref{tilde-Omega-x}--\eqref{tilde-Q} with \eqref{rescaled-GP} and \eqref{direct_transformation_A-PD_parameters-alpha=0}, the metric functions in the GP metric \eqref{ds2_accel_kerr_new_polar} become
\begin{align}
 \tilde{\Omega} & = 1\,,  \nonumber\\
 \tilde{\rho}^{\,2}   &= \tilde{r}^{\,2}+\frac{1}{a^2} \big(l +a\,\cos\theta\big)^2\,, \nonumber\\
 \tilde{P} &= 1 \,,\nonumber\\
 \tilde{Q} &= \frac{1}{a^2} \big((a^2-l^2+e^2+g^2) - 2m\,a\,\tilde{r} +a^2\tilde{r}^{\,2} \big). \label{tilde-Q-Aalpha=0}
\end{align}
This gives an explicit form of the metric in terms of the \emph{Astorino physical parameters}. Restoring the correct dimensionality of the coordinates and parameters by the choice
\begin{equation}\label{gamma_for_a=0}
    \gamma = a \,,
\end{equation}
we finally obtain
\begin{align}
 \tilde{\rho}^{\,2}   &= \tilde{r}^{\,2}+(l +a\,\cos\theta)^2 \,, \nonumber\\
 \tilde{Q}  &= (a^2-l^2+e^2+g^2)-2m\,\tilde{r}+\tilde{r}^{\,2} \,, \nonumber
\end{align}
which fully agrees with Eq.~(16.23) in Sec.~16.3.1 of~\cite{GriffithsPodolsky:2009}.

\newpage

\subsection {The special case ${a=0}$: no Kerr-like rotation}
\label{sc:a=0}

In such a case the auxiliary parameters are
\begin{align}
    I& = 1-\alpha^2l^2, \qquad
    a|J| = 2\alpha l^2, \qquad
    aK = \alpha l^2, \qquad \nonumber\\
    aL&= \alpha\,(2m l+e^2+g^2), \qquad
    aM = \alpha^2 l\, (-2l^2+e^2+g^2),
\end{align}
so that using \eqref{direct_transformation_A-PD-acceleration-explicit-again}, \eqref{a_tld}, \eqref{l_tld}, \eqref{omega_0} we obtain
\begin{align}
    \tilde{\alpha}&= \alpha^2l^2, \nonumber\\[2mm]
    \tilde{a}&= 2\sqrt{\omega_0}\,\dfrac{1}{1-\alpha^2l^2},\qquad
    \tilde{l}= \dfrac{\sqrt{\omega_0}}{\alpha l}\,\dfrac{1+\alpha^2l^2}{1-\alpha^2l^2}, \label{omega_0-for-a=0}\\[2mm]
    \omega_0&= \frac{1-\alpha^2l^2}{ 1 - 2\alpha^2 ml + \alpha^4l^2(e^2+g^2-l^2) }. \nonumber
\end{align}
Consequently,
\begin{align}
 \tilde{P}& = 1-a_3\cos\theta-a_4\cos^2\theta,   \label{aux-0}\\
 \tilde{Q}& = \dfrac{\omega_0^2}{4\alpha^2l^4}
    \Big[(1-\alpha^2l^2) - 2\alpha l (1+\alpha^2l^2)\dfrac{\tilde{r}}{\sqrt{\omega_0}}
      +\alpha^2l^2 (1-\alpha^2l^2)\dfrac{\tilde{r}^{\,2}}{\omega_0}\,\Big]  \nonumber\\
    &\hspace{11mm}\times\Big[e^2+g^2-2ml - 2\alpha l\,(e^2+g^2-2l^2)\dfrac{\tilde{r}}{\sqrt{\omega_0}}
            +\alpha^2l^2(e^2+g^2+2ml)\dfrac{\tilde{r}^{\,2}}{\omega_0}\,\Big], \label{aux-1}
\end{align}
where
\begin{align}
    a_3 & = 2\alpha\,
    \frac{ m\,(1+\alpha^2l^2) - l - \alpha^2 l \,(e^2+g^2-l^2)}
    {1 - 2\alpha^2 ml + \alpha^4l^2(e^2+g^2-l^2)},\\[2mm]
    a_4 & = \alpha^2\dfrac{2ml-e^2-g^2}{ 1 - 2\alpha^2 ml + \alpha^4l^2(e^2+g^2-l^2)}. \label{aux-2}
\end{align}
Here we have used the definition \eqref{tild_P_simpl-again} together with \eqref{mathcalP} \eqref{mathcalQ}, that is
\begin{eqnarray}
\tilde{\mathcal{P}}(\tilde{x}) :=
   \dfrac{\omega_0^2}{\tilde{a}^2}\,
   P'\Big(\sqrt{\omega_0}\,x'=\dfrac{\tilde{l}+\tilde{a}\,\tilde{x}}{\sqrt{\omega_0}}\Big), \qquad
\tilde{\mathcal{Q}}(\tilde{r}) :=
   \omega_0^2\,
   Q' \Big(r'=\dfrac{\tilde{r}}{\sqrt{\omega_0}}\Big),
\end{eqnarray}
in which the explicit transformation \eqref{coord_tr_2},
\begin{equation}
    x' = \dfrac{\tilde{l}+\tilde{a}\,\tilde{x}}{\omega_0}
       = \frac{(1+\alpha^2l^2) + 2\alpha l\,\tilde{x}}{\alpha l\,(1-\alpha^2l^2)\,\sqrt{\omega_0}},
\end{equation}
was inserted. In fact, the resulting metric functions $\tilde{P}$, $\tilde{Q}$ are fully consistent with the expressions \eqref{P'-factorized:a=0},  \eqref{Q'-factorized:a=0}. In particular, the first quadratic factor in $P'$ gives the term ${(1-\tilde{x}^{\,2})}$, while the second term leads to ${\tilde{P}(\tilde{x}) = 1-a_3\,\tilde{x}-a_4\,\tilde{x}^2}$. It is then natural to introduce ${\tilde{x}=\cos\theta}$.

Correct dimensionality of the physical parameters can be restored by applying \eqref{resc} with
\begin{equation}\label{gamma_for_a=0-repeated}
    \gamma = \alpha\, l^2,
\end{equation}
so that
\begin{align}
    \tilde{\alpha}&= \alpha,\qquad
    \tilde{a}= 2\sqrt{\omega_0}\,\dfrac{\alpha\, l^2}{1-\alpha^2l^2},\qquad
    \tilde{l}= l\,\sqrt{\omega_0}\,\dfrac{1+\alpha^2l^2}{1-\alpha^2l^2}.
         \label{omega_0-for-a=0rescaled}
\end{align}
Recall that the mass and charge parameters $m,e,g$ in \eqref{aux-1}--\eqref{aux-2} already have the proper physical dimension (of length) but it is also necessary to rescale ${\tilde{r} \mapsto \gamma\,\tilde{r}}$, ${ \tilde{Q} \mapsto \gamma^2 \,\tilde{Q}}$. Thus,
\begin{align}
 \tilde{Q}(\tilde{r}) &  = \dfrac{1}{4}\,
    \Big[(1-\alpha^2l^2)\,\omega_0 - 2(1+\alpha^2l^2)\sqrt{\omega_0}\,\dfrac{\tilde{r}}{l}
      +(1-\alpha^2l^2)\,\dfrac{\tilde{r}^{\,2}}{l^2 }\,\Big]  \nonumber\\
    &\hspace{4mm}\times
    \Big[(e^2+g^2-2ml)\,\omega_0 - 2(e^2+g^2-2l^2)\sqrt{\omega_0}\,\dfrac{\tilde{r}}{l}
      +(e^2+g^2+2ml)\,\dfrac{\tilde{r}^{\,2}}{l^2 }\,\Big], \label{aux-3}
\end{align}
while the function $\tilde{P}$ remains the same as in \eqref{aux-0}. Also, here we keep the same dimensionless parameter $\omega_0$ given by \eqref{omega_0-for-a=0}.

We can thus write the metric for \emph{accelerating charged NUT black hole without the Kerr-like rotation in the Griffiths-Podolsk\'y metric} representation~\eqref{ds2_accel_kerr_new_polar} as
\begin{align}
    \dd s^2=\dfrac{1}{{\tilde{\Omega}}^2}
    \Bigg[&-\dfrac{\tilde{Q}}{\,\tilde{\rho}^{\,2}}
    \Big[\,
    \dd t - \big( (\tilde{a}+\tilde{l}) + (\tilde{l}+\tilde{a}\,\cos\theta)\big)(1-\cos\theta)\,\dd\varphi
    \,\Big]^2
     +\dfrac{\,\tilde{\rho}^{\,2}}{\tilde{Q}}\,\dd\tilde{r}^2\nonumber\\
    &+\dfrac{\,\tilde{\rho}^{\,2}}{\tilde{P}}\,\dd\theta^2
     +\dfrac{\tilde{P}}{\,\tilde{\rho}^{\,2}}\sin^2\theta\,
    \big[\,\tilde{a}\,\dd t-\big(\tilde{r}^2+(\tilde{a}+\tilde{l})^2\big)\,\dd\varphi\big]^2\Bigg],
    \label{ds2_accel_NUT}
\end{align}
where from \eqref{omega_0-for-a=0rescaled} it follows that
\begin{align}
  \tilde{a}+\tilde{l}             &= l\,\sqrt{\omega_0}\,\,\dfrac{1+\alpha\,l}{1-\alpha\,l}, \label{auxili-7}\\
  \tilde{l}+\tilde{a}\,\cos\theta &= l\,\sqrt{\omega_0}\,\Big[\,\dfrac{1+\alpha\,l}{1-\alpha\,l} - \dfrac{2\alpha\,l}{1-\alpha^2l^2}(1-\cos\theta) \Big]. \label{auxili-8}
\end{align}
Applying these relations we get an explicit metric
\begin{align}
    \dd s^2=\dfrac{1}{{\tilde{\Omega}}^2}
    \Bigg\{&\!\!-\dfrac{\tilde{Q}}{\,\tilde{\rho}^{\,2}}
    \Big[\,
    \dd t -
    2l\,\sqrt{\omega_0}\,\Big[\,\dfrac{1+\alpha\,l}{1-\alpha\,l} - \dfrac{\alpha\,l}{1-\alpha^2l^2}(1-\cos\theta) \Big]
    (1-\cos\theta)\,\dd\varphi
    \,\Big]^2
     +\dfrac{\,\tilde{\rho}^{\,2}}{\tilde{Q}}\,\dd\tilde{r}^2\nonumber\\
    &+\dfrac{\,\tilde{\rho}^{\,2}}{\tilde{P}}\,\dd\theta^2
     +\dfrac{\tilde{P}}{\,\tilde{\rho}^{\,2}}\sin^2\theta\,
    \Big[\,2\sqrt{\omega_0}\,\dfrac{\alpha\, l^2}{1-\alpha^2l^2}\,\dd t
    -\Big(\tilde{r}^2+l^2\omega_0\,\dfrac{(1+\alpha\,l)^2}{(1-\alpha\,l)^2}\Big)\,\dd\varphi\Big]^2\Bigg\},
    \label{ds2_accel_NUT-explicit}
\end{align}
in which \eqref{tilde-Omega-omega-gamma} and \eqref{tilde-rho} takes the form
\begin{align}
  \tilde{\Omega}       &= 1-\Big[\,\dfrac{1+\alpha\,l}{1-\alpha\,l} - \dfrac{2\alpha\,l}{1-\alpha^2l^2}(1-\cos\theta) \Big]
  \,\dfrac{\tilde{r}}{l \sqrt{\omega_0}}, \label{tilde-Omega-a=0}\\
  \tilde{\rho}^{\,2}   &= \tilde{r}^{\,2}+l^2\omega_0 \Big[\,\dfrac{1+\alpha\,l}{1-\alpha\,l} - \dfrac{2\alpha\,l}{1-\alpha^2l^2}(1-\cos\theta) \Big]^2, \label{tilde-rho-a=0}
\end{align}
while the functions $\tilde{P}$ an  $\tilde{Q}$ are given by \eqref{aux-0} and \eqref{aux-3}, respectively.

It seems convenient now to perform a rescaling of the coordinates and the metric functions
\begin{align}
  \bar{r}  &= \frac{\tilde{r}}{\sqrt{\omega_0}}, \qquad
  \bar{t}   = \sqrt{\omega_0}\,t, \qquad
  \bar{\varphi} = \omega_0\,\varphi,
   \label{resca1e1-a=0}\\
  \bar{P}  &= \frac{\tilde{P}}{\omega_0}, \qquad
  \bar{Q}   = \frac{\tilde{Q}}{\omega_0^2}, \qquad
  \bar{\rho}^{\,2} = \frac{\tilde{\rho}^{\,2}}{\omega_0},
  \label{resca1e-2a=0}
\end{align}
which brings the metric to the form
\begin{align}
    \dd s^2=\dfrac{1}{{\bar{\Omega}}^2}
    \Bigg\{&\!\!-\dfrac{\bar{Q}}{\,\bar{\rho}^{\,2}}
    \Big[\,
    \dd \bar{t} - 2l\,\frac{ (1+\alpha^2l^2)(1-\cos\theta) + \alpha\,l \sin^2\theta }{1-\alpha^2l^2}\,    \dd\bar{\varphi} \,\Big]^2
     +\dfrac{\,\bar{\rho}^{\,2}}{\bar{Q}}\,\dd\bar{r}^2\nonumber\\
    &+\dfrac{\,\bar{\rho}^{\,2}}{\bar{P}}\,\dd\theta^2
     +\dfrac{\bar{P}}{\,\bar{\rho}^{\,2}}\sin^2\theta\,
    \Big[\,\dfrac{2\,\alpha\, l^2}{1-\alpha^2l^2}\,\dd \bar{t}
    -\Big(\bar{r}^2+l^2\,\dfrac{(1+\alpha\,l)^2}{(1-\alpha\,l)^2}\Big)\dd\bar{\varphi}\,\Big]^2\Bigg\},
    \label{ds2_accel_NUT-rescaled}
\end{align}
where
\begin{align}
  \bar{\Omega}    &= 1-
  \dfrac{1+\alpha^2l^2+2\alpha\,l\cos\theta}{1-\alpha^2l^2} \,\dfrac{\bar{r}}{l}, \label{bar-Omega-a=0}\\[2mm]
  \bar{\rho}^{\,2}&= \bar{r}^{\,2}+l^2
   \Big[\,\dfrac{1+\alpha^2l^2+2\alpha\,l\cos\theta}{1-\alpha^2l^2}\,\Big]^2 , \label{bar-rho-a=0}\\[2mm]
  \bar{P}         &= \frac{1}{1-\alpha^2l^2}\Big[\,{1 - 2\alpha^2 ml + \alpha^4l^2(e^2+g^2-l^2)} \nonumber\\
      &\hspace{22mm} + 2\alpha \big(\, l - m\,(1+\alpha^2l^2) + \alpha^2 l \,(e^2+g^2-l^2)\big)\cos\theta  \nonumber\\
      &\hspace{22mm} + \alpha^2 ( e^2+g^2-2ml)\cos^2\theta \Big],\label{bar-P-a=0}\\[2mm]
  \bar{Q}         &= \frac{1}{4l^4}\Big[\,(\bar{r}-l)^2 - \alpha^2 l^2(\bar{r}+l)^2\Big]
       \Big[\,2ml(\bar{r}^2-l^2) + 4l^3\bar{r} +(e^2+g^2)(\bar{r}-l)^2 \Big].\label{bar-Q-a=0}
\end{align}
This is an explicit GP metric form~\eqref{ds2_accel_kerr_new_polar} of the class of \emph{accelerating charged NUT black holes} (of type~D) \emph{without the Kerr-like rotation}. It has not been identified in previous works \cite{GriffithsPodolsky:2005, GriffithsPodolsky:2006, PodolskyGriffiths:2006, PodolskyVratny:2021, PodolskyVratny:2023} due to the fact that the Kerr-like parameter $\tilde{a}$ was not properly chosen to cover this special subcase. Its convenient choice is \eqref{omega_0-for-a=0rescaled}, coupled to the NUT parameter~$l$ and acceleration~$\alpha$, whereas previously it was incorrectly assumed that ${\tilde{a}=0}$ in such a situation. This lead to a degenerate parametrization of this sector of the complete family of type D black hole spacetimes.

In this metric we can \emph{independently} set ${\alpha=0}$ and ${l=0}$, expecting to obtain the (charged) NUT solution without acceleration and the (charged) C-metric without the NUT parameter, respectively. Let us investigate these two special subcases in detail.\\

For ${\alpha=0}$ we get
\begin{align}
    \dd s^2=
    & -\dfrac{\bar{Q}}{{\bar{\Omega}}^2\,\bar{\rho}^{\,2}}
    \big[\, \dd \bar{t} -  2l(1-\cos\theta)\dd\bar{\varphi}  \,\big]^2
     +\dfrac{\,\bar{\rho}^{\,2}}{{\bar{\Omega}}^2\bar{Q}}\,\dd\bar{r}^2
     +\frac{\bar{r}^2+l^2}{{\bar{\Omega}}^2} \big(\,\dd\theta^2  + \sin^2\theta\,\dd\bar{\varphi}^2\,\big),
    \label{ds2_NUT-rescaled-alpha=0}
\end{align}
with ${\bar{P} = 1}$ and
\begin{align}
  \bar{\rho}^{\,2}&= \bar{r}^{\,2}+l^2, \label{bar-rho-a=0-alpha=0}\\
  \bar{\Omega}    &= \dfrac{1}{l} (l-\bar{r}), \label{bar-Omega-a=0-alpha=0}\\
  \bar{Q}         &= \frac{1}{4l^4}\,(l-\bar{r})^2
       \Big[\,2ml(\bar{r}^2-l^2) + 4l^3\bar{r} +(e^2+g^2)(\bar{r}-l)^2 \Big].\label{bar-Q-a=0-alpha=0}
\end{align}
Performing now the following transformation and rescaling of the physical parameters,
\begin{align}
  \frac{\bar{r}}{l}   &= \frac{R-L}{R+L},\qquad
  \bar{t} = \sqrt{2}\,T,        \label{transf-to-standatd-NUT1}\\
  l &=\sqrt{2}\,L,\qquad
  m=\sqrt{2}\,M,\qquad
  e=\sqrt{2}\,E,\qquad
  g=\sqrt{2}\,G,  \label{transf-to-standatd-NUT2}
\end{align}
we obtain
\begin{align}
    \dd s^2=
    -F  \big[\,\dd T - 2L(1-\cos\theta)\dd\bar{\varphi} \,\big]^2
     +\dfrac{\dd R^2}{F} + (R^2+L^2)(\,\dd\theta^2  + \sin^2\theta\,\dd\bar{\varphi}^2),
    \label{ds2_NUT-standard}
\end{align}
where
\begin{align}
  F = \frac{R^2-2MR-L^2+E^2+G^2}{R^2+L^2}.\label{F_NUT-standard}
\end{align}
This is the \emph{standard form of the NUT solution}, see Eqs.~(12.3), (12.2) and~(12.19) in \cite{GriffithsPodolsky:2009}. For vanishing NUT parameter (${L=0}$), it reduces to the Reissner-Nordstr\"om solution.

It should also be remarked that the inverse transformation to \eqref{transf-to-standatd-NUT1} reads
\begin{align}
 \frac{R}{L} = \frac{l+\bar{r}}{l-\bar{r}},
\end{align}
so that ${R=\infty \Leftrightarrow \bar{r}=l \Leftrightarrow \bar{\Omega}=0 }$. The conformal infinity is thus approached as ${R\to\infty}$ in the standard coordinates of \eqref{ds2_NUT-standard}, while it is located at ${\bar{r}=l}$ in the new (unfamiliar) metric representation \eqref{ds2_accel_NUT-rescaled} when ${\alpha=0}$. This may be another ``technical'' reason why it was previously difficult to identify the genuine accelerating \emph{purely NUT} black holes in the whole type D class.\\

\newpage

Finally, let us investigate the complementary special subcase ${l=0}$, which should lead to the (charged) C-metric without the NUT parameter. To do so, we consider the metric \eqref{ds2_accel_NUT-rescaled}--\eqref{bar-Q-a=0} and perform the change of coordinates
\begin{equation}
       \bar{r}  = \frac{l\, R}{R+2l}\,, \qquad   \bar{\varphi} = \varphi + \alpha\, \bar{t} \,.   \label{transf-to-standatd-C-metric1}
\end{equation}
It is then possible in the transformed line element to take the limit ${l \to 0}$ to zero NUT parameter, obtaining
\begin{align}
    2\,\dd s^2 = \dfrac{1}{(1-\alpha\,R\cos\theta)^2}
    \Big[ - \bar{Q}\, \dd \bar{t}^{\,2} + \dfrac{\dd R^2}{\bar{Q}}
     +\dfrac{R^{2}}{\bar{P}}\,\dd\theta^2
     +\bar{P}\,R^2\sin^2\theta\, \dd \varphi^2\, \Big],
     \label{ds2_C-metric-standard}
\end{align}
where
\begin{align}
  \bar{P} &= 1 - 2\alpha m\,\cos\theta  + \alpha^2 ( e^2+g^2)\cos^2\theta \,,  \label{bar-P-a=0l=0} \\
  \bar{Q} &= \big( 1 - \alpha^2R^2\big)\Big(1 - \frac{2m}{R} + \frac{e^2+g^2}{R^2} \Big) \,. \label{bar-Q-a=0l=0}
\end{align}
This is precisely the \emph{standard form of the C-metric solution}, as given by Eqs.~(14.6), (14.41) in \cite{GriffithsPodolsky:2009} (up to a trivial overall conformal rescaling of the line element $\dd s^2$ by the constant factor~2). Recall that the roots of $Q$ identify horizons. In general, there are two \emph{acceleration horizons} located at ${R_{{\rm a}\pm} = \pm 1/\alpha}$ and two \emph{black-hole horizons} at ${R_{{\rm b}\pm} = \tilde{m} \pm \sqrt{\tilde{m}^{\,2}-\tilde{e}^{\,2}-\tilde{g}^{\,2}}}$ \cite{GriffithsPodolsky:2005, GriffithsPodolsky:2006, PodolskyGriffiths:2006, PodolskyVratny:2021}. For vanishing acceleration (${\alpha=0}$), the metric \eqref{ds2_C-metric-standard} reduces to the usual form of the Reissner-Nordstr\"om solution.

\newpage

\section{Transformation to the Podolsk\'y-Vr\'atn\'y metric form}
\label{sc:PV-form}

Finally, we will elucidate the relation between the Astorino (A) metric \eqref{init_metr} of all type~D black holes and the metric presented by Podolsk\'y and Vr\'atn\'y (PV) in 2021 and 2023 \cite{PodolskyVratny:2021, PodolskyVratny:2023}. The PV metric is an improvement of the GP metric in the sense that the key metric functions are considerably simpler, fully explicit and factorized. This turned out to be convenient for geometrical and physical interpretation of these spacetimes (identification and description of horizons and singularities, finding the global structure, ergoregions, identification of cosmic string, including regions with closed timelike curves if these strings are rotating, etc.).

Actually, the new PV metric has the \emph{same general form} as the GP metric \eqref{ds2_accel_kerr_new_polar}, that is the metric given by Eqs.~(6.18) in~\cite{GriffithsPodolsky:2009}, but the metric functions $\tilde{P}, \tilde{Q}$ are much simpler  (this was achieved by introducing a new set of the mass and charge parameters, rescaling the metric by a uniquely chosen constant conformal factor, and making a suitable choice of the twist parameter $\omega$). Therefore, we can employ the same initial steps as in section~\ref{sc:GP-form}, starting form the PD$_{\alpha}$ form of the metric \eqref{PD-GP-form-again} --- that is equivalent to the A form --- arriving at  \eqref{ds2_accel_kerr_new}--\eqref{mathcalQ}. Changing tildes to hats in all the parameters and coordinates (except~$t$ and~$\varphi$ which remain the same), we thus obtain
\begin{align}
    \dd s^2=\dfrac{1}{{\hat{\Omega}}^2}
    \Bigg[&-\dfrac{\hat{\mathcal{Q}}}{\,\hat{\rho}^{\,2}}
    \Big[\dd t-\big(\hat{a}(1-\hat{x}^2)+2\hat{l}(1-\hat{x})\big)\dd\varphi\Big]^2
     +\dfrac{\,\hat{\rho}^{\,2}}{\hat{\mathcal{Q}}}\,\dd\hat{r}^2\nonumber\\
    &+\dfrac{\,\hat{\rho}^{\,2}}{\hat{\mathcal{P}}}\,\dd\hat{x}^2
     +\dfrac{\hat{\mathcal{P}}}{\,\hat{\rho}^{\,2}}
    \big[\hat{a}\,\dd t-\big(\hat{r}^2+(\hat{a}+\hat{l})^2\big)\,\dd\varphi\big]^2\Bigg],
    \label{ds2_accel_kerr_new-PV}
\end{align}
with
\begin{eqnarray}
\hat{\Omega} :=& 1-\dfrac{\alpha'}{\omega}\,\hat{r}\,(\hat{l}+\hat{a}\,\hat{x}),\label{omega-PV} \\[2mm]
\hat{\rho}^{\,2} :=& \hat{r}^{\,2}+(\hat{l}+\hat{a}\,\hat{x})^2, \\[2mm]
\hat{\mathcal{P}}(\hat{x}) :=& \dfrac{\omega^2}{\hat{a}^2}\,P'\Big(\dfrac{\hat{l}+\hat{a}\,\hat{x}}{\sqrt{\omega}}\Big),\\[1mm]
\hat{\mathcal{Q}}(\hat{r}) :=& \hspace{-4mm} \omega^2\, Q' \Big(\dfrac{\hat{r}}{\sqrt{\omega}}\Big)\,,
\label{mathcalQ-PV}
\end{eqnarray}
where the functions $P'(x'), Q'(r')$ are given by \eqref{P'-factorized}, \eqref{Q'-factorized}. Again, this is the general Griffiths-Podolsk\'y metric, see Eq.~(16.12) in~\cite{GriffithsPodolsky:2009}.\\

The conformal factor $\hat{\Omega}$ already has the PV form, so that we can directly identify the Podolsk\'y-Vr\'atn\'y (dimensionless) \emph{acceleration parameter} as
\begin{equation}\label{hat-alpha}
    \hat{\alpha} = \tilde{\alpha} = \alpha' = \alpha\,a\,\big[\,K - \alpha^2(a^2-l^2)\big].
\end{equation}
The PV, GP, and PD acceleration parameters are thus \emph{the same}, and expressed explicitly in terms on the Astorino parameters,
\begin{equation}\label{direct_transformation_A-PD-acceleration-explicit-again2}
 \hat{\alpha} = \alpha\,\,\tfrac{1}{2}\big[\,a - \alpha^2 a (a^2-l^2)
+ \sqrt{a^2+\alpha^4 a^2(a^2-l^2)^2+2\alpha^2(a^2-l^2)(a^2-2l^2)}\, \big].
\end{equation}
If ${\alpha=0}$ then ${\hat{\alpha}=0}$. As in the GP case, for ${l=0}$ we get ${\hat{\alpha}=\alpha\,a}$, while for ${a=0}$ we get ${ \hat{\alpha}=\alpha^2 l^2}$.

Concerning the PV \emph{rotational parameters} $\hat{a}$ and $\hat{l}$, as in the previous GP case they are fixed by the condition that the metric function $\hat{\mathcal{P}}(\tilde{x})$ is factorized to ${\hat{\mathcal{P}} \propto (1-\hat{x}^2)}$, this leads to the same expressions as in \eqref{a_tld} and \eqref{l_tld}, that is
\begin{align}
    \hat{a} = \tilde{a}&= \dfrac{ \sqrt{\omega}}{K-\alpha^2(a^2-l^2)}\,
              \frac{\sqrt{I^2 \mp J^2}}{I},\label{a_tld-PV}\\
    \hat{l} = \tilde{l}&= \dfrac{ \sqrt{\omega}}{K-\alpha^2(a^2-l^2)}\,
              F\,\dfrac{l}{a}\,. \label{l_tld-PV}
\end{align}
Recall that the upper (minus) sign in \eqref{a_tld-PV} applies when ${a^2>l^2}$, while the lower (plus) sign is used when ${a^2<l^2}$. With these parameters, the metric function $\hat{\mathcal{P}}$ in \eqref{ds2_accel_kerr_new-PV} takes the form analogous to \eqref{tild_P_x},
\begin{align}
 \hat{\mathcal{P}}(\hat{x})=&\ (1-\hat{x}^2)\,\dfrac{\omega^2}{\hat{a}^2}\,\dfrac{1}{I\,[K-\alpha^2(a^2-l^2)]}\,\nonumber\\
    & \times \Bigg[\,L  - 2M\dfrac{\hat{l}}{\sqrt{\omega}}
       +\alpha^2\big[(a^2-l^2)L+(e^2+g^2)\,\sqrt{I^2 \mp J^2}\,\big] \dfrac{\hat{l}^2}{\omega}\nonumber\\
    &\qquad
       -\Big(2M\dfrac{\hat{a}}{\sqrt{\omega}}+2\alpha^2\big[(a^2-l^2)L+(e^2+g^2)\,\sqrt{I^2 \mp J^2}\,\big]
       \dfrac{\hat{l}\hat{a}}{\omega}\Big)\,\hat{x}\nonumber\\
    &\qquad
       +\alpha^2\,\dfrac{\hat{a}^2}{\omega}\big[(a^2-l^2)L+(e^2+g^2)\,\sqrt{I^2 \mp J^2}\,\big]\,\hat{x}^2 \Bigg].\label{tild_P_x-PV}
\end{align}

So far, the steps were the same as those deriving the GP form of the metric in previous section. However now, instead of fixing the twist parameter  ${\omega=\omega_0}$ by \eqref{omega_0}, we choose a \emph{different value} ${\omega=\omega_1}$, where
\begin{align}
    \omega_1 = \dfrac{\alpha\, a}{\hat{\alpha}}\dfrac{I^2 \mp J^2}{I\,L}
      \equiv \dfrac{I^2 \mp J^2}{I \big[\,K - \alpha^2(a^2-l^2)\big] L }
  \,.\label{omega_1}
\end{align}
The function $\hat{\mathcal{P}}$ then becomes
\begin{align}
    \hat{\mathcal{P}}=&(1-\hat{x}^2)\Bigg[1-2\dfrac{M}{L}\dfrac{\hat{l}+\hat{a}\,\hat{x}}{\sqrt{\omega_1}}
    +\alpha^2\Big[a^2-l^2+(e^2+g^2)\,\dfrac{\sqrt{I^2 \mp J^2}}{L}\,\Big]\Big(\dfrac{\hat{l}+\hat{a}\,\hat{x}}{\sqrt{\omega_1}}\Big)^2\Bigg].
\end{align}

Now we introduce \emph{new mass and charge parameters} as
\begin{align}
    \hat{m}& := \dfrac{1}{\hat{\alpha}}\,\dfrac{M}{L}\,\sqrt{\omega_1}\,,\label{hat_m}\\
    \hat{e}^2+\hat{g}^2& :=\dfrac{\alpha^2}{\hat{\alpha}^2}\,(e^2+g^2)\,\dfrac{\sqrt{I^2 \mp J^2}}{L}\,\omega_1\,,\label{hat_eg}
\end{align}
and employ an important identity
\begin{equation}\label{identity_alpha^2(a^2-l^2)}
    \hat{\alpha}^2(\hat{a}^2-\hat{l}^{\,2})=\alpha^2(a^2-l^2)\,\omega_1\,,
\end{equation}
so that $\hat{\mathcal{P}}$ becomes

\begin{equation}\label{hat-mathcal-P}
\hat{\mathcal{P}}=(1-\hat{x}^2)\Bigg[1-2\,\hat{\alpha}\,\hat{m}\,\dfrac{\hat{l}+\hat{a}\,\hat{x}}{\omega_1}
    +\hat{\alpha}^2\big(\hat{a}^2-\hat{l}^2+\hat{e}^2+\hat{g}^2\big)\Big(\dfrac{\hat{l}+\hat{a}\hat{x}}{\omega_1}\Big)^2\Bigg].
\end{equation}
Defining
\begin{equation}\label{r+-}
    \hat{r}_{\pm} :=\hat{m}\pm\sqrt{\hat{m}^2+\hat{l}^2-\hat{a}^2-\hat{e}^2-\hat{g}^2},
\end{equation}
the metric function takes a \emph{compact and fully factorized form}
\begin{equation}\label{facorized-hat-mathcal-P}
    \hat{\mathcal{P}}(\hat{x})=(1-\hat{x}^2)\Big(1-\dfrac{\hat{\alpha}}{\omega_1}\,\hat{r}_{+}(\hat{l}+\hat{a}\,\hat{x})\Big)
         \Big(1-\dfrac{\hat{\alpha}}{\omega_1}\,\hat{r}_{-}(\hat{l}+\hat{a}\,\hat{x})\Big).
\end{equation}
Notice that such a factorization corresponds to the factorization of the Astorino metric function \eqref{delta_x_init},
\begin{equation}\label{Delta_x-factorized}
  \Delta_x = (1-x^2)\big( 1 - \alpha\, r_+\, x \big) \big( 1 - \alpha\, r_-\, x\big),
\end{equation}
where ${r_{\pm} := m \pm \sqrt{m^2+l^2-a^2-e^2-g^2}}$.\\

Similarly we  analyze the metric function $\hat{\mathcal{Q}}$ in \eqref{ds2_accel_kerr_new-PV}. From the definition \eqref{mathcalQ-PV} with \eqref{Q'-factorized}, we obtain the expression
\begin{align}
 \hat{\mathcal{Q}}(\hat{r})=\dfrac{\omega_1^2}{I^2 \mp J^2}&
    \Bigg[I-2\alpha l[1-\alpha^2(a^2-l^2)]\,\dfrac{\hat{r}}{\sqrt{\omega_1}}
    -\alpha^2(a^2-l^2)I\,\dfrac{\hat{r}^{\,2}}{\omega_1} \Bigg]\nonumber\\
    \times& \Bigg[ \dfrac{K-1}{\alpha^2(a^2-l^2)^2}\,\big[(a^2-l^2)L + (e^2+g^2)\sqrt{I^2 \mp J^2}\,\big] \nonumber
     \\& -\dfrac{2}{\alpha a}M\,\dfrac{\hat{r}}{\sqrt{\omega_1}}+L\,[K-\alpha^2(a^2-l^2)]\,\dfrac{\hat{r}^{\,2}}{\omega_1}\Bigg]. \label{hat-mathcal-Q'}
\end{align}
Using the relation ${\hat{\alpha}\,\hat{l} = \tilde{\alpha}\,\tilde{l} = \alpha\,l\,\sqrt{\omega_1}\, F}$ which follows form  \eqref{l_tld-alter}, and the identity \eqref{identity_alpha^2(a^2-l^2)}, the first (square) bracket can be rewritten in the factorized form
\begin{align}
 I\,\Big(1-2\,\hat{\alpha}\,\hat{l}\,\dfrac{\hat{r}}{\omega_1}-\hat{\alpha}^2(\hat{a}^2-\hat{l}^{\,2})\dfrac{\hat{r}^2}{\omega_1^2}\Big)
=I\,\Big(1-\hat{\alpha}(\hat{a}+\hat{l}\,)\dfrac{\hat{r}}{\omega_1}\Big)
 \Big(1+\hat{\alpha}(\hat{a}-\hat{l}\,)\dfrac{\hat{r}}{\omega_1}\Big).
\end{align}
The second (square) bracket, applying the relations
\begin{equation}
    \dfrac{K-1}{\alpha^2(a^2-l^2)^2} = \dfrac{\alpha}{\hat{\alpha}\,a}\,,\qquad
    K-\alpha^2(a^2-l^2) = \dfrac{\hat{\alpha}}{\alpha\,a}\,,
\end{equation}
see \eqref{hat-alpha}, and the definitions of the ``hatted'' charges and masses \eqref{hat_m} and \eqref{hat_eg}, becomes
\begin{align}
    \frac{\hat{\alpha}}{\alpha\,a}\,L\,
    \Big[\dfrac{\alpha^2(a^2-l^2)}{\hat{\alpha}^2} + (\hat{e}^2+\hat{g}^2)\dfrac{1}{\omega_1}
    - 2 \,\hat{m}\,\dfrac{\hat{r}}{\omega_1} + \dfrac{\hat{r}^2}{\omega_1} \Big].
\end{align}
Using the identity \eqref{identity_alpha^2(a^2-l^2)} we then obtain a factorized expression
\begin{align}
   \dfrac{\hat{\alpha}}{\alpha\,a}\,\dfrac{L}{\omega_1}\,
   \big[\hat{r}^2-2\,\hat{m}\,\hat{r}+(\hat{a}^2-\hat{l}^2+\hat{e}^2+\hat{g}^2)\big]
  = \dfrac{\hat{\alpha}}{\alpha\,a}\,\dfrac{L}{\omega_1}\,(\hat{r}-\hat{r}_+)(\hat{r}-\hat{r}_-).
\end{align}
Therefore, the metric function  \eqref{hat-mathcal-Q'} takes the form
\begin{equation}
\hat{\mathcal{Q}}= \omega_1\,\dfrac{\hat{\alpha}}{\alpha\,a}\,\dfrac{I\,L}{I^2 \mp J^2}
 \,(\hat{r}-\hat{r}_+)(\hat{r}-\hat{r}_-)
 \Big(1-\hat{\alpha}(\hat{a}+\hat{l}\,)\dfrac{\hat{r}}{\omega_1}\Big)
 \Big(1+\hat{\alpha}(\hat{a}-\hat{l}\,)\dfrac{\hat{r}}{\omega_1}\Big),
\end{equation}
which for the convenient choice \eqref{omega_1} of $\omega_1$ simplifies to a \emph{nice factorized expression}
\begin{equation}\label{facorized-hat-mathcal-Q}
    \hat{\mathcal{Q}}(\hat{r})=
    (\hat{r}-\hat{r}_+)(\hat{r}-\hat{r}_-)
    \Big(1-\hat{\alpha}(\hat{a}+\hat{l}\,)\dfrac{\hat{r}}{\omega_1}\Big)    \Big(1+\hat{\alpha}(\hat{a}-\hat{l}\,)\dfrac{\hat{r}}{\omega_1}\Big).
\end{equation}
This directly corresponds to the factorization of the Astorino metric function \eqref{delta_r_init},
\begin{equation}\label{Delta_r-factorized}
  \Delta_r = (r-r_+)(r-r_-)(1-\alpha\,r)(1+\alpha\,r)\,.
\end{equation}

Finally, using \eqref{hat-alpha} the conformal factor \eqref{omega-PV} reads
${\hat{\Omega} = 1-\dfrac{\hat{\alpha}}{\omega_1}\,\hat{r}\,(\hat{l}+\hat{a}\,\hat{x})}$. As argued in section~\ref{sc:issue-twist}, the special (nonzero) twist parameter $\omega_1$ can by put to any (nonzero) value $\omega$ by the simple rescaling \eqref{rescaling-of-alpha} of the acceleration parameter (provided it is nonzero),
\begin{equation}\label{rescaling-of-alpha-to-omega}
    \hat{\alpha} \quad\mapsto\quad \hat{\alpha}\,\dfrac{\omega_1}{\omega}\,.
\end{equation}
It explicitly \emph{restores a freedom to choose}~$\omega$ in $\hat{\Omega}$, and also in
the metric functions \eqref{facorized-hat-mathcal-P} and \eqref{facorized-hat-mathcal-Q}. The metric  \eqref{ds2_accel_kerr_new-PV} thus becomes
\begin{align}
    \dd s^2=\dfrac{1}{{\hat{\Omega}}^2}
    \Bigg[&-\dfrac{\hat{\mathcal{Q}}}{\,\hat{\rho}^{\,2}}
    \Big[\dd t-\big(\hat{a}(1-\hat{x}^2)+2\hat{l}(1-\hat{x})\big)\dd\varphi\Big]^2
     +\dfrac{\,\hat{\rho}^{\,2}}{\hat{\mathcal{Q}}}\,\dd\hat{r}^2\nonumber\\
    &+\dfrac{\,\hat{\rho}^{\,2}}{\hat{\mathcal{P}}}\,\dd\hat{x}^2
     +\dfrac{\hat{\mathcal{P}}}{\,\hat{\rho}^{\,2}}
    \big[\hat{a}\,\dd t-\big(\hat{r}^2+(\hat{a}+\hat{l})^2\big)\,\dd\varphi\big]^2\Bigg],
    \label{ds2_accel_kerr_new-PV-x}
\end{align}
with
\begin{align}
\hat{\Omega} &= 1-\dfrac{\hat{\alpha}}{\omega}\,\hat{r}\,(\hat{l}+\hat{a}\,\hat{x}),\label{omega-PV2} \\[2mm]
\hat{\rho}^{\,2} &= \hat{r}^{\,2}+(\hat{l}+\hat{a}\,\hat{x})^2, \\[2mm]
\hat{\mathcal{P}}(\hat{x}) &=(1-\hat{x}^2)\Big(1-\dfrac{\hat{\alpha}}{\omega}\,\hat{r}_{+}(\hat{l}+\hat{a}\,\hat{x})\Big)
         \Big(1-\dfrac{\hat{\alpha}}{\omega}\,\hat{r}_{-}(\hat{l}+\hat{a}\,\hat{x})\Big),\\[1mm]
\hat{\mathcal{Q}}(\hat{r}) &=   (\hat{r}-\hat{r}_+)(\hat{r}-\hat{r}_-)
    \Big(1-\hat{\alpha}(\hat{a}+\hat{l}\,)\dfrac{\hat{r}}{\omega}\Big)    \Big(1+\hat{\alpha}(\hat{a}-\hat{l}\,)\dfrac{\hat{r}}{\omega}\Big)\,.
\label{mathcalQ-PV-x}
\end{align}

It  now only remains to introduce the angular coordinate via the relation ${\hat{x}=\cos\theta}$, and  \emph{choose} the twist parameter to be
\begin{equation}\label{PV-choice-of-omega}
    \omega :=\dfrac{\hat{a}^2+\hat{l}^{\,2}}{\hat{a}}\,.
\end{equation}
This finally leads to
\begin{align}
    \dd s^2=\dfrac{1}{{\hat{\Omega}}^2}
    \Bigg[&-\dfrac{\hat{Q}}{\,\hat{\rho}^{\,2}}
    \Big[\dd t-\big(\hat{a}\sin^2\theta+2\hat{l}(1-\cos\theta)\big)\dd\varphi\Big]^2
     +\dfrac{\,\hat{\rho}^{\,2}}{\hat{Q}}\,\dd\hat{r}^2\nonumber\\
    &+\dfrac{\,\hat{\rho}^{\,2}}{\hat{P}}\,\dd\theta^2
     +\dfrac{\hat{P}}{\,\hat{\rho}^{\,2}}\sin^2\theta\,
    \big[\,\hat{a}\,\dd t-\big(\hat{r}^2+(\hat{a}+\hat{l})^2\big)\,\dd\varphi\big]^2\Bigg],
    \label{PV-metric}
\end{align}
where
\begin{align}
\hat{\Omega} &= 1-\dfrac{\hat{\alpha}\,\hat{a}}{\hat{a}^2+\hat{l}^{\,2}}
\,\hat{r}\,(\hat{l}+\hat{a}\cos\theta)\,, \label{PV-hat-Omega} \\[2mm]
\hat{\rho}^{\,2} &= \hat{r}^{\,2}+(\hat{l}+\hat{a}\,\cos\theta)^2, \\[2mm]
\hat{P}(\theta) &=\Big(1-\dfrac{\hat{\alpha}\,\hat{a}}{\hat{a}^2+\hat{l}^{\,2}}
                         \,\hat{r}_{+}(\hat{l}+\hat{a}\cos\theta)\Big)
                  \Big(1-\dfrac{\hat{\alpha}\,\hat{a}}{\hat{a}^2+\hat{l}^{\,2}}
                         \,\hat{r}_{-}(\hat{l}+\hat{a}\cos\theta)\Big),\\[1mm]
\hat{Q}(\hat{r}) &=   (\hat{r}-\hat{r}_+)(\hat{r}-\hat{r}_-)
    \Big(1-\hat{\alpha}\,\hat{a}\,\dfrac{\hat{a}+\hat{l}}{\hat{a}^2+\hat{l}^{\,2}}\,\hat{r}\Big)    \Big(1+\hat{\alpha}\,\hat{a}\,\dfrac{\hat{a}-\hat{l}}{\hat{a}^2+\hat{l}^{\,2}}\,\hat{r}\Big). \label{PV-hat-Q}
\end{align}
This is exactly the Podolsk\'y-Vr\'atn\'y form of the metric, as expressed by  Eqs.~(1)--(5) in~\cite{PodolskyVratny:2021}.

Let us emphasize that the choice \eqref{PV-choice-of-omega}, although convenient in many situations, prevents us to identify accelerating ``purely NUT'' black holes because ${\hat{a}=0}$ completely removes the acceleration parameter $\hat{\alpha}$ from the metric \eqref{PV-metric}--\eqref{PV-hat-Q}. However, because $\omega$ is an additional free parameter, the PV metrics \eqref{ds2_accel_kerr_new-PV-x} with \emph{different values of the twist~$\omega$ are equivalent} by the rescaling \eqref{rescaling-of-alpha-to-omega} of the acceleration parameter (unless ${\omega=0}$). This keeps a possibility to \emph{explicitly identify the accelerating purely NUT black holes} (without the Kerr-like rotation, charged, of type~D) even within the PV class of metrics for a better choice of $\omega$ than \eqref{PV-choice-of-omega}, such that the degeneracy ${\hat{a}=0 \Rightarrow \hat{\alpha}=0}$ is removed. This possibility was explicitly demonstrated in the GP form of the metric, see Eqs.~\eqref{ds2_accel_NUT-rescaled}--\eqref{bar-Q-a=0}.

\section{Summary of the special cases}
\label{sc:special-cases}

Let us summarize special cases of the Astorino (A, or A$^+$) metric in which the physical parameters are set to zero, and the corresponding Pleba\'nski-Demia\'nski (PD), Griffiths-Podolsk\'y (GP) and Podolsk\'y-Vratn\'y (PV)  parameters.

\subsection {The limit ${l \to 0}$}
As derived in \eqref{direct_transformation_A-PD_parameters-l=0}, the dimensionless PD acceleration parameter is ${\alpha' = \alpha\, a}$. After restoring the correct physical dimensionality of the GP and PV parameters in Subsection~\ref{sc:l=0}, the result is simple, namely
\begin{align}
\tilde{\alpha} & = \hat{\alpha} = \alpha\,, \nonumber\\
\tilde{a}  & =\hat{a}=a\,,  \\
\tilde{l}  & =\hat{l}=l=0\,. \nonumber
\end{align}
The acceleration, Kerr-like and NUT parameters in A, A$^+$, GP, and PV metric forms are thus \emph{identical}.

\subsection{The limit ${\alpha \to 0}$}

From \eqref{direct_transformation_A-PD-acceleration-explicit} we obtain
\begin{equation}
    \alpha' \to \alpha\, a,\qquad\hbox{so that}\qquad
    \alpha' = \tilde{\alpha} = \hat{\alpha} \to 0.
\end{equation}
The Kerr-like and NUT parameters (rescaled to a correct physical dimension) are
\begin{align}
\tilde{a} &=\hat{a}=a\,,\nonumber\\
\tilde{l} &=\hat{l}=l\,,
\end{align}
respectively. This gives us the Kerr-Newman-NUT solution in A, A$^+$, GP, and PV metrics.

\subsection{The limit ${a \to 0}$}

From \eqref{a_tld}--\eqref{def-F} it follows that
\begin{equation}
    \dfrac{\tilde{a}}{\tilde{l}} = \frac{\sqrt{I^2 \mp J^2}}{1-\alpha^2(a^2-l^2)}\,\frac{a}{l}
    \quad\to\quad  \dfrac{2 \alpha l}{1+\alpha^2 l^2}.
\end{equation}
In the limit ${a \to 0}$ the ratio ${\tilde{a}/\tilde{l}}$ thus \emph{remains finite, and nonzero} (unless ${\alpha l=0}$). It means that taking the limit in which Astorino's Kerr-like parameter $a$ tends to zero \emph{does not} bring either of the PV parameters $\tilde{a}$ or $\tilde{l}$ to zero.

Thus we conclude that the accelerating \emph{purely} NUT black hole solution with ${a=0}$, identified by Astorino, corresponds (after a suitable  coordinate transformations presented in previous sections) to \emph{``accelerating Kerr-NUT solution''} with a \emph{non-zero} Kerr-like GP parameter ${\tilde{a}\ne0}$ (and ${\tilde{l}\ne0}$), and PV parameter ${\hat{a}\ne0}$ (and ${\hat{l}\ne0}$).

This also explains why in the A (and A$^+$) metric form the acceleration  parameter $\alpha$ is \emph{not redundant} in the case ${a=0}$, as opposed in the GP (and PV) form. Similar considerations show that the GP, PV, and PD \emph{acceleration parameters are non-zero} in the case ${a=0}$. Indeed, from \eqref{hat-alpha}, \eqref{direct_transformation_A-PD-acceleration-explicit-again2} we obtain that the \emph{dimensionless} acceleration parameters are given by
\begin{equation}\label{hat-alpha-a=0}
    \tilde{\alpha} = \hat{\alpha} = \alpha' = \alpha^2 l^2\,,
\end{equation}
which does \emph{not} vanish when ${\alpha l \ne 0}$.


\section{Conclusions}
\label{sc:conclusions}

We thoroughly studied a large class of spacetimes representing black holes with mass $m$, rotation~$a$, NUT parameter~$l$, acceleration~$\alpha$, and electric and magnetic charges $e$ and $g$ (generating electromagnetic field aligned with both principal null directions of the type D Weyl tensor).

In particular, we found relations between coordinates and physical parameters of various metric forms of such exact solutions to the Einstein-Maxwell system, namely those of Pleba\'nski-Demia\'nski (PD), Astorino (A, improved here to a more compact metric A$^+$), Griffiths-Podolsk\'y (GP), and Podolsk\'y-Vr\'atn\'y (PV). The references to original articles, nomenclature and conventions are summarized in Table~\ref{Tab-summary-of-metrics}.\\

Main conclusions resulting from our investigation are:

\begin{itemize}

\item If properly mapped and physically interpreted, all these representations cover the complete class of such type D black holes.

\item The physical parameters of the A metric representation (and thus also A$^+$) are a very good choice for describing the type D black holes. Moreover, they can be set to zero in any order, leading to expected special cases (and further subcases), without any unpleasant coordinate degeneracies or divergencies.

\item Explicit coordinate transformations and relations between the parameters of A, A$^+$, PD, GP and PV metric forms (and their variants which also include the twist parameter $\omega$) were found and discussed. These mutual relations are shown in the scheme on Figure~\ref{Fig-scheme} by arrows, with the references to the corresponding Sections of our paper.

\item In the subcase of vacuum black holes, the relation between A and PD representations is in agreement with the results of \cite{WuWu:2024}.

\item We clarified the role of the twist parameter $\omega$, related to both the Kerr-like rotation parameter~$a$ and to the NUT parameter~$l$.

\item Expressed in terms of the new convenient physical parameters $m, a, l, \alpha, e, g$, the key A$^+$, PD, GP and PV metric functions are explicitly factorized into the product of quadratic expressions. This is very helpful for the physical interpretation, namely the identification of horizons and axes.

\item Special attention was payed to the main subclasses of type D black holes, namely those with no NUT (${l=0}$), no acceleration (${\alpha=0}$) and --- until recently elusive --- black holes with no Kerr like-rotation ${a=0}$.

\item We proved that in the subclasses ${l=0}$ and ${\alpha=0}$, the rotation, NUT, and acceleration parameters are the same in  A, GP, and PV metric forms, that is ${a=\tilde{a}=\hat{a}}$, ${l=\tilde{l}=\hat{l}}$, and ${\alpha=\tilde{\alpha}= \hat{\alpha}}$.

\item On the other hand, in the non-rotating subclass ${a=0}$, the Kerr-like parameters ${\tilde{a}=\hat{a}}$ and the NUT parameters ${\tilde{l}=\hat{l}}$ were not properly identified in previous GP and PV metric forms \cite{GriffithsPodolsky:2005, GriffithsPodolsky:2006, PodolskyGriffiths:2006, PodolskyVratny:2021, PodolskyVratny:2023}. The correct general relations are given by \eqref{a_tld}--\eqref{def-F}, which in the case ${a=0}$ reduce to \eqref{omega_0-for-a=0rescaled}.

\item The accelerating (charged) purely NUT black holes of type D (${a=0}$, but $\alpha l$ nonzero) thus have ${\tilde{a}=\hat{a}\ne0}$ in the GP and PV forms. With this correction, we were able to derive the GP (equivalent to PV) metric form of such black holes \eqref{ds2_accel_NUT-rescaled}--\eqref{bar-Q-a=0}.

\item In this metric it is possible to independently set ${\alpha=0}$ and ${l=0}$, obtaining thus the (charged) NUT solution without acceleration and the (charged) C-metric without the NUT parameter, respectively, in their usual form.

\end{itemize}

\section*{Acknowledgments}

This work has been supported by the Czech Science Foundation Grant No.~GA\v{C}R 23-05914S.

\vspace{5mm}

\newpage

\appendix

\section{Derivation of the transformation to the Pleba\'nski-Demia\'nski form of the metric, and identification of the physical parameters}
\label{systematic-derivation}

The  A$^+$ metric \eqref{ds2_simpl} has a general form which is very similar to the Griffiths--Podolsk\'y representation~\cite{GriffithsPodolsky:2005, GriffithsPodolsky:2006, PodolskyGriffiths:2006} (see also Eq.~(16.18) in \cite{GriffithsPodolsky:2009}, and ~\cite{PodolskyVratny:2021, PodolskyVratny:2023})
of the family of type D black holes in the Pleba\'nski-Demia\'nski class of electrovacuum solutions with~$\Lambda$.
Nevertheless, a closer look at both the metrics reveals a \emph{crucial difference}. In the Griffiths--Podolsk\'y representation, the functions $A(x)$ and $C(r)$ are  \emph{constants}, namely
\begin{align} \label{Ac-Cc}
    A \equiv A_c = 1,
       \qquad
    C \equiv C_c = a
\end{align}
(there is a nice agreement with \eqref{Aw}, \eqref{Cw} for ${\alpha=0}$).
However, we can remedy this problem by a suitable \emph{transformation of the coordinates} ${x \mapsto X}$ and ${r \mapsto R}$ given by
\begin{align} \label{Xx-Rr}
    \dd X \equiv A_c\,\dfrac{\dd x}{A(x)},
        \qquad
    \dd R \equiv C_c\,\dfrac{\dd r}{C(r)}.
\end{align}
We also introduce new metric functions, expressed in these coordinates, as
\begin{align} \label{Deltax-DeltaR}
    \Delta_X(X) \equiv \dfrac{A_c^2}{A^2}\,\Delta_x,
        \qquad
    \Delta_R(R) \equiv \dfrac{C_c^2}{C^2}\,\Delta_r,
\end{align}
and analogously
\begin{align} \label{Omega-rho}
    {\tilde\Omega}^2(R,X) \equiv \dfrac{A_c C_c}{A\,C}\,\Omega^2,
        \qquad
    {\tilde\rho}^{\,2}(R,X)   \equiv \dfrac{A_c C_c}{A\,C}\,\rho^2,
\end{align}
so that
\begin{align} \label{Omega-rho-invar}
    \dfrac{{\tilde\rho}^{\,2}}{\tilde\Omega^2} = \frac{\rho^2}{\Omega^2}.
\end{align}
Then the A$^+$ metric \eqref{ds2_simpl} takes the form
\begin{equation}
    \dd s^2=\dfrac{1}{{\tilde\Omega}^2}\bigg[
    - \dfrac{\Delta_R}{{\tilde\rho}^{\,2}}(A_c\,\dd t - \mathcal{B}\,\dd\varphi)^2
    + \dfrac{\Delta_X}{{\tilde\rho}^{\,2}}(C_c\,\dd t + \mathcal{D}\,\dd\varphi)^2
    + C_f\,{\tilde\rho}^{\,2} \Big(\,\dfrac{\dd R^2}{\Delta_R} + \dfrac{\dd X^2}{\Delta_X}\,\Big)\bigg]
    \label{ds2_simpl-new},
\end{equation}
where $A_c, C_c$ are constants, as desired, and the new metric functions are
\begin{align} \label{calB-calD}
    \mathcal{B}(X) \equiv A_c \dfrac{B}{A},
        \qquad
    \mathcal{D}(R) \equiv C_c \dfrac{D}{C}.
\end{align}

It is now possible to find an explicit relation between the Astorino representation of \emph{all type D black holes}, written in the form (\ref{ds2_simpl-new}), and the Pleba\'nski-Demia\'nski class of the same solutions, written in the Griffiths--Podolsk\'y representation. In particular, this will elucidate the problem of \emph{purely NUT accelerating black holes} of type D without the Kerr-like rotation (i.e., black holes with ${m, l, \alpha \ne 0}$ and ${a=0}$, admitting also the charges ${e, g}$ and a cosmological constant $\Lambda$). These are clearly contained in the Astorino class of solutions (\ref{init_metr}) but \emph{so far have not been identified} in the Pleba\'nski-Demia\'nski family.

The procedure starts with an integration of (\ref{Xx-Rr}),
\begin{align}\label{integration-of-X-R}
    X=A_c\int\dfrac{\dd x}{A(x)},
       \qquad
    R=C_c\int\dfrac{\dd r}{C(r)},
\end{align}
of the specific quadratic functions $A(x)$ and $C(r)$ given by (\ref{Aw}) and (\ref{Cw}), respectively. Fortunately, it can be done explicitly, yielding for ${a^2>l^2}$ quite a simple transformation
\begin{align}
 x =&\  \dfrac{1}{\alpha\sqrt{a^2-l^2}}\,
 \tanh\Big[\dfrac{\alpha\sqrt{a^2-l^2}}{A_c}\,(X + X_0) \Big], \label{def-x-and-r-1}\\
 r =&\  \dfrac{\sqrt{a^2-l^2}}{\alpha a}\,
 \tan\Big[\dfrac{\alpha\sqrt{a^2-l^2}}{C_c}\, (R + R_0) \Big] - \dfrac{l}{\alpha a},
   \label{def-x-and-r}
\end{align}
where $X_0, R_0$ are free constants of integration. Notice that for very small values of $X$ and $R$ we obtain just linear relations
${ x \approx \dfrac{1}{A_c}\,(X + X_0)}$ and
${ r + \dfrac{l}{\alpha a} \approx \dfrac{a^2-l^2}{C_c\,a}\,(R + R_0)}$.

Here \emph{we assume} ${a^2>l^2}$, i.e., that the Kerr-like rotation parameter $a$ is greater than the NUT parameter $l$, and ${a>0}$. The complementary case ${a^2<l^2}$ can be treated similarly by changing the trigonometric functions to hyperbolic ones, and vice versa. It leads to the \emph{same expressions}, but with the term  ${\sqrt{a^2-l^2}}$ generalized to ${\sqrt{|a^2-l^2|}}$, and ${2K := I + \sqrt{I^2+J^2}}$ instead of the definition ${2K := I + \sqrt{I^2-J^2}}$ employed in \eqref{defI-and -defJ-repeated}. It also covers various special cases when (some of) parameters are zero.

Next we have to evaluate the functions  $A(X), B(X), C(R), D(R)$ given by \eqref{Aw}--\eqref{Bu}. Then we obtain $\mathcal{B}(X), \mathcal{D}(R)$ by using \eqref{calB-calD}, and finally
$\Delta_X(X), \Delta_R(R), {\tilde\Omega}^2(R,X), {\tilde\rho}^{\,2}(R,X)$  by using \eqref{Deltax-DeltaR}, \eqref{Omega-rho}.

A direct substitution into \eqref{Aw}, \eqref{Cw} gives
\begin{align} \label{A(X)-C(R)}
    \frac{1}{A(X)} = \cosh^2\xi ,
       \qquad
    \frac{1}{C(R)} = \dfrac{a}{a^2-l^2}\,\cos^2\chi,
\end{align}
where we introduced convenient dimensionless parameters
\begin{align} \label{def-x-and-r-with X0-and-R_0}
     \xi \equiv \dfrac{\alpha\,\sqrt{a^2-l^2}}{A_c}\,(X + X_0) \equiv \xi' + \xi_0\, ,
       \qquad
    \chi \equiv \dfrac{\alpha\,\sqrt{a^2-l^2}}{C_c}\,(R + R_0) \equiv \chi' + \chi_0\,.
\end{align}
General expressions for $B(X), D(R)$  look more complicated, namely
\begin{align}
B(X) =&\ \dfrac{a}{\alpha^2(a^2-l^2)}\tanh^2 \xi + \dfrac{2l}{\alpha\sqrt{a^2-l^2}}\tanh \xi + a \,,
\label{B_expr} \\
D(R) = &\ \dfrac{1}{\alpha^2a^2}\Big[(a^2-l^2)\tan^2\chi - 2l\sqrt{a^2-l^2}\,\tan \chi + [l^2+\alpha^2 a^2(l^2-a^2) ] \Big]\,,\label{D_expr}
\end{align}
and thus using \eqref{calB-calD}
\begin{align}
    \mathcal{B}(X) = &\  \dfrac{A_c\,a}{\alpha^2(a^2-l^2)}\Big[
    \sinh \xi \,\big(I \sinh \xi + J \cosh \xi \big) + \alpha^2(a^2-l^2)
    \Big],\label{m_expr} \\[2mm]
    \mathcal{D}(R) = &\  \dfrac{C_c}{\alpha^2(a^2-l^2)a}\Big[
    \sin \chi\,\big( (a^2 I - 2 l^2) \sin\chi - 2l\sqrt{a^2-l^2} \cos\chi\Big)
    +l^2-\alpha^2(a^2-l^2)a^2\Big],\label{p_expr}
\end{align}
where the constants $I$, $J$, and $K$ are defined as
\begin{align} \label{defI-and -defJ-repeated}
    I  &:= 1+\alpha^2(a^2-l^2)\, , \nonumber\\
    J  &:= 2 l\,\dfrac{\alpha}{a}\,\sqrt{a^2-l^2}\,,\\
    2K &:= I + \sqrt{I^2-J^2}\,. \nonumber
\end{align}

Following the idea employed previously in \cite{DebeverKamranMcLenaghan:1984} (see Section~4 in and subcases A2 and A3 therein), these involved functions can be considerably simplified by employing a remaining coordinate freedom encoded in the integration constants $X_0, R_0$ of \eqref{A(X)-C(R)}. By their \emph{unique choice}
\begin{align}
    \tanh\big(2 \xi_0\big) = &\ -\frac{J}{I},\label{choice-of-X0} \\[2mm]
    \tan \big(2 \chi_0\big)= &\
       \dfrac{2l\sqrt{a^2-l^2}}{a^2 I - 2 l^2}
       \qquad
       \equiv  \dfrac{a}{\alpha}\,\dfrac{J}{a^2 I - 2 l^2},
    \label{choice-of-R0}
\end{align}
(assuming ${a^2 I > 2 l^2}$, which is natural because it admits a special case ${l=0}$) we achieve
\begin{align}
    \mathcal{B}(X) = &\  A_c\, (\, b_0 + b_1\, \sinh^2 \xi' \,),\label{m_expr2} \\[2mm]
    \mathcal{D}(R) = &\  C_c\, (\, d_0 + d_1\, \sin^2 \chi' \,),\label{p_expr2}
\end{align}
where
the constant coefficients read
\begin{align}
    b_1 = +\,d_1 = &\  \frac{a}{\alpha^2(a^2-l^2)}\,\sqrt{I^2-J^2},\label{def-b_0} \\[2mm]
    b_0 = -\,d_0 = &\  \frac{a}{\alpha^2(a^2-l^2)}\,(K - 1).\label{def-b_1}
\end{align}

Next step is to apply a transformation of all the coordinates
\begin{align}
    t=&\ \dfrac{1}{2} \Big[\alpha\sqrt{a^2-l^2}\,(\sqrt{I^2-J^2}-I+2)\,\tau'+(\sqrt{I^2-J^2}+I-2)\,\frac{\phi'}{\alpha\sqrt{a^2-l^2}}\,\Big],
    \label{transf-t}\\
    \varphi=&\
    \dfrac{1}{a}
    \Big[\alpha\sqrt{a^2-l^2}\,\phi'-\big(\alpha\sqrt{a^2-l^2}\,\big)^3\,\tau'\Big],
    \label{transf-phi}\\
    \tanh \xi'=&\ \alpha\, \sqrt{a^2-l^2}\,x' ,\label{transf-xi'}\\
    \tan \chi'=&\ \alpha\, \sqrt{a^2-l^2}\,r' ,\label{transf-chi'}
\end{align}
so that
\begin{align}    \label{miscal-1}
(\dd t - b_0\,\dd\varphi) = (\dd t + d_0\,\dd\varphi) = \alpha\sqrt{a^2-l^2}\,\sqrt{I^2-J^2}\,\,\dd\tau' \, ,
\end{align}
and
\begin{align}    \label{miscal-2}
b_1 \sinh^2 \xi'  = \frac{a \sqrt{I^2-J^2}\,{x'}^2}{1-\alpha^2(a^2-l^2)\, {x'}^2} , \qquad
d_1 \sin^2 \chi'  = \frac{a \sqrt{I^2-J^2}\,{r'}^2}{1+\alpha^2(a^2-l^2)\, {r'}^2} .
\end{align}
This leads to a great simplification of the key combinations of the functions
\begin{align}    \label{key-terms-in-ds2_simpl-new}
A_c\,\big(\dd t - (\, b_0 + b_1 \sinh^2 \xi' \,)\,\dd\varphi\big) = &\
  A_c\,\frac{\alpha\sqrt{a^2-l^2}\sqrt{I^2-J^2}} {1-\alpha^2(a^2-l^2)\, {x'}^2} \,\big(\,\dd\tau'  -  {x'}^2\, \dd\phi' \,\big) ,\\
C_c\,\big(\dd t + (\, d_0 + d_1 \sin^2 \chi' \,)\,\dd\varphi\big) = &\
  C_c\,\frac{\alpha\sqrt{a^2-l^2}\sqrt{I^2-J^2}} {1+\alpha^2(a^2-l^2)\, {r'}^2} \,\big(\,\dd\tau'  +  {r'}^2\, \dd\phi' \,\big).
\end{align}

Using \eqref{Omega-rho}, \eqref{rho2def}, \eqref{calB-calD}  we also obtain ${{\tilde\rho}^{\,2}=A_c\,\mathcal{D}+C_c\,\mathcal{B}}$. The functions $\mathcal{D}, \mathcal{B}$ are given by \eqref{p_expr2}, \eqref{m_expr2}, \eqref{miscal-2}, so that (with the help of the relation ${d_0+b_0=0}$)
\begin{align} \label{tilde-rho-appendix}
    {\tilde\rho}^{\,2} = \frac{A_c C_c\,a \,\sqrt{I^2-J^2}\,({r'}^2+{x'}^2)}
    {\big[1+\alpha^2(a^2-l^2)\, {r'}^2\big]\big[1-\alpha^2(a^2-l^2)\, {x'}^2\big]} .
\end{align}

If we now conveniently introduce metric functions $\Omega'(r', x')$, $P(x')$, $Q(r')$ as
\begin{align}
\Omega'^{\,2} \equiv &\ a\,\frac{\big[1+\alpha^2(a^2-l^2)\, {r'}^2\big]
\big[1-\alpha^2(a^2-l^2)\, {x'}^2\big]}{A_c C_c\,\alpha^2(a^2-l^2) \sqrt{I^2-J^2}}\,\,{\tilde\Omega}^2 , \label{Omega-definition}\\[1mm]
P' \equiv &\ A_c^{-2}\, [1-\alpha^2(a^2-l^2)\, {x'}^2]^2\,\Delta_X , \label{P-definition}\\[3mm]
Q' \equiv &\ C_c^{-2}\, [1+\alpha^2(a^2-l^2)\, {r'}^2]^2\,\Delta_R , \label{Q-definition}
\end{align}
the metric \eqref{ds2_simpl-new}  takes a nice form
\begin{align}     \label{ds2-new-in-PD-form}
    \dd s^2=\dfrac{1}{\Omega'^{\,2}}\bigg[
    &- \dfrac{Q'}{{r'}^2+{x'}^2}\,\big(\,\dd\tau'  -  {x'}^2\, \dd\phi' \,\big)^2
     + \dfrac{P'}{{r'}^2+{x'}^2}\,\big(\,\dd\tau'  +  {r'}^2\, \dd\phi' \,\big)^2 \\
    &+ \frac{a^2\,C_f}{\alpha^2(a^2-l^2)}\,({r'}^2+{x'}^2)
    \Big(\,\dfrac{\dd {r'}^2}{Q'} + \dfrac{\dd {x'}^2}{P'}\,\Big)\bigg].\nonumber
\end{align}
Actually, this is a general Pleba\'nski-Demia\'nski metric representing all (double aligned, non-null) solutions of Einstein-Maxwell-$\Lambda$ equations of algebraic type~D, including black holes of this type, see Eq.~(3.30) in~\cite{PlebanskiDemianski:1976}, and Chapter~16 of~\cite{GriffithsPodolsky:2009} in which a special gauge ${\alpha^2(a^2-l^2)=-1}$ and ${a^2\,C_f=-1}$ is considered. The former condition can always be obtained by a rescaling on the angular coordinate~$\phi'$, while the latter is achieved by the choice of the free constant $C_f$. Also, these operations relate the metrics \eqref{ds2-new-in-PD-form} and \eqref{PD-GP-form}.

\newpage
To complete the transformation from the Astorino new metric representation \eqref{init_metr}-\eqref{delta_x_init} to the Pleba\'nski-Demia\'nski metric representation \eqref{ds2-new-in-PD-form}, it remains to prove that the \emph{conformal factor} \eqref{Omega-definition} has the form ${\Omega' = 1-\alpha'\,r'x' }$, with a suitable acceleration parameter $\alpha'$, and also that the metric functions $P'(x')$, $Q'(r')$ are \emph{quartic} functions of the respective coordinates.

Moreover, this explicit transformation will yield a \emph{unique relation between the physical parameters of the Astorino metric and the (purely mathematical) Pleba\'nski-Demia\'nski  parameters}, and with the physical parameters in the Griffiths-Podolsk\'y form of these black-hole metrics. In particular, it will give their fully general form, which will identify the elusive (overlooked) accelerating NUT black holes without the Kerr-like rotation parameter~$a$.

To this end, we have to explicitly evaluate these metric functions. We start with the conformal factor ${\tilde\Omega}^2$ given by \eqref{Omega-rho}, \eqref{Om_cf},
\begin{align}\label{Omega-1}
   {\tilde\Omega}^2=&\ \frac{A_c C_c}{A\,C}\, (1-\alpha\, r\, x)^2.
\end{align}
In view of \eqref{def-x-and-r}, \eqref{A(X)-C(R)}, \eqref{def-x-and-r-with X0-and-R_0} we get
\begin{align}\label{Omega-2}
{\tilde\Omega}^2 =  A_c C_c\,\dfrac{a}{a^2-l^2}\, \Big[ \cos\chi\, \cosh\xi
 - \dfrac{1}{\alpha\sqrt{a^2-l^2}} \,\sin (\chi - \beta) \sinh\xi \,\Big]^2,
\end{align}
where we defined a useful auxiliary constant $\beta$ as
\begin{equation}
\sin\beta \equiv \frac{l}{a} ,
\qquad\hbox{so that}\quad
\cos\beta = \frac{\sqrt{a^2-l^2}}{a},
\quad\hbox{and}\quad
\tan\beta = \frac{l}{\sqrt{a^2-l^2}}.
\end{equation}
Recalling \eqref{def-x-and-r-with X0-and-R_0}
 we rewrite \eqref{Omega-2} as
\begin{align}\label{Omega-3}
{\tilde\Omega}^2 =  A_c C_c\,\dfrac{a}{\alpha^2(a^2-l^2)^2} \, \Big[& -\alpha\sqrt{a^2-l^2}\,\cos(\chi' + \chi_0)\, \cosh(\xi' + \xi_0) \nonumber\\
 &\quad +\sin \big(\chi' + (\chi_0 - \beta)\big) \sinh(\xi' + \xi_0) \,\Big]^2,
\end{align}
where
\begin{align}
    \tanh\big(2 \xi_0\big) = -\frac{J}{I},
       \qquad
    \tan \big(2 \chi_0\big) = \dfrac{a}{\alpha}\,\dfrac{J}{a^2 I - 2 l^2}
    \label{choice-of-R0-repeated}
\end{align}
see  \eqref{choice-of-X0}, \eqref{choice-of-R0}. Using standard identities for goniometric and hyperbolic functions, one derives the equivalent expressions (to be employed below), namely
\begin{align}
    \cosh\big(2 \xi_0\big) = \frac{I}{\sqrt{I^2-J^2}},
       \qquad
    \cos \big(2 \chi_0\big) = \frac{a^2 I - 2 l^2}{a^2\,\sqrt{I^2-J^2}},
    \label{choice-of-R0-B}
\end{align}
so that
\begin{align}
    \cosh^2\xi_0 = \frac{I+\sqrt{I^2-J^2}}{2\sqrt{I^2-J^2}},
      \qquad
    \cos^2\chi_0 =&\  \frac{a^2(\sqrt{I^2-J^2}+I) - 2l^2}{2a^2\,\sqrt{I^2-J^2}},
    \label{choice-of-R0-C} \\
    \sinh^2\xi_0 = \frac{I-\sqrt{I^2-J^2}}{2\sqrt{I^2-J^2}},
      \qquad
    \sin^2\chi_0 =&\  \frac{a^2(\sqrt{I^2-J^2}-I) + 2l^2}{2a^2\,\sqrt{I^2-J^2}},
    \label{choice-of-R0-D}
\end{align}
and thus
\begin{align}
    \tanh \xi_0 =&\  -\frac{J}{2K},     \label{choice-of-R0-E1}\\
    \tan \chi_0 =&\  \frac{a}{\alpha}\,\frac{J}{2(a^2 K-l^2)}.
    \label{choice-of-R0-E}
\end{align}
Notice that ${l=0}$ implies ${J=0=\beta}$, so that ${\xi_0 = 0 = \chi_0}$.

Applying these relations, it is possible to prove that
\begin{align}\label{Omega-identities}
  \alpha\sqrt{a^2-l^2}\, \sin\chi_0 \cosh \xi_0 = &\ -\cos(\chi_0-\beta) \sinh \xi_0,\\
  \alpha\sqrt{a^2-l^2}\, \cos\chi_0 \sinh \xi_0 = &\ +\sin(\chi_0-\beta) \cosh \xi_0.
\end{align}
Using these two identities, after employing in \eqref{Omega-3} usual formulae for the sum in the argument of trigonometric and hyperbolic function , the terms containing ${\cos\chi' \sinh \xi'}$ and ${\sin\chi' \cosh \xi'}$ vanish, while the two remaining combine in such a way that
\begin{align}\label{Omega-4}
{\tilde\Omega}^2 =  A_c C_c\,\dfrac{a}{a^2-l^2} \, \Big[\,
 \frac{\cos\chi_0}{\cosh\xi_0}\,\cos\chi' \cosh\xi'
+\frac{\sin\chi_0}{\sinh\xi_0}\,\sin\chi' \sinh\xi' \,\Big]^2.
\end{align}
Performing now the transformation \eqref{transf-xi'}, \eqref{transf-chi'} we get
\begin{align}\label{Omega-5}
{\tilde\Omega}^2 =  A_c C_c\,\dfrac{a\,c_0^2 }{a^2-l^2} \,
\frac{\big(\,1 - \alpha'\,r'\,x'\,\big)^2}
{\big[1+\alpha^2(a^2-l^2)\, {r'}^2\big]\big[1-\alpha^2(a^2-l^2)\, {x'}^2\big]},
\end{align}
where
\begin{align}\label{c_0^2-alpha'}
c_0^2 =  \frac{\cos^2\chi_0}{\cosh^2\xi_0}, \qquad
\alpha' = \alpha^2(a^2-l^2)\,\frac{\tan\chi_0}{\tanh \xi_0} .
\end{align}
The conformal factor \eqref{Omega-definition} thus, using \eqref{choice-of-R0-C} and \eqref{choice-of-R0-E},
takes the form
\begin{align}\label{conformal-relation}
\Omega'^{\,2} = S^2 \big(\,1 - \alpha'\,r'\,x'\,\big)^2,
\end{align}
where the dimensionless parameters are
\begin{align}
\alpha' &=\   \alpha\,a\,\frac{(a^2-l^2)K}{a^2K-l^2}, \label{alpha'}\\
S^2     &=\   \frac{1}{\alpha^2(a^2-l^2)^2}\,\frac{a^2K-l^2}{K \sqrt{I^2-J^2}}
\quad \equiv \dfrac{a^2}{\alpha^4(a^2-l^2)^3}\dfrac{K-1}{\sqrt{I^2-J^2}}. \label{S^2}
\end{align}

For ${l=0}$, these relations simplify considerably to  ${\alpha' =  \alpha\,a}$ and ${S^{-2}=\alpha^2 a^2 (1+\alpha^2a^2)}$. On the contrary, for ${a=0}$, it reduces to ${\alpha' = \alpha^2\,l^2}$.

Using these relations, we can rewrite the metric (\ref{ds2-new-in-PD-form}) in the form:
\begin{align}
    \dd s^2=\dfrac{1}{S^2(1-\alpha'r'x')^2}\bigg[
    &- \dfrac{Q}{{r'}^2+{x'}^2}\,\big(\,\dd\tau'  -  {x'}^2\, \dd\phi' \,\big)^2 \nonumber\\
    &+ \dfrac{P}{{r'}^2+{x'}^2}\,\big(\,\dd\tau'  +  {r'}^2\, \dd\phi' \,\big)^2 \\
    &+ \frac{a^2\,C_f}{\alpha^2(a^2-l^2)}\,({r'}^2+{x'}^2) \Big(\,\dfrac{\dd {r'}^2}{Q} + \dfrac{\dd {x'}^2}{P}\,\Big)\bigg].\nonumber
\end{align}
By rescaling time and angular coordinates
\begin{equation}
    \tau' \mapsto S\,\tau'\,,\qquad  \phi' \mapsto S\,\phi'\,,
\end{equation}
and introducing a new constant
\begin{equation}
    c^2 :=\dfrac{a^2 C_f}{\alpha^2(a^2-l^2)S^2}\,,
\end{equation}
the metric takes exactly the form \eqref{PD-GP-form}. Thus, to conclude, the required transformation of coordinates from the A to the PD form of the spacetimes is given by
\begin{align}
t &=\ \dfrac{a}{\alpha(a^2-l^2)}\,\sqrt{\dfrac{K-1}{\sqrt{I^2-J^2}}}\,
    \Big[\big[K-\alpha^2(a^2-l^2)\big]\,\tau' + \frac{K-1}{\alpha^2(a^2-l^2)}\,\phi'\,\Big],\label{t-trans-fin}\\
\varphi &=\ \dfrac{1}{\alpha(a^2-l^2)}\sqrt{\dfrac{K-1}{\sqrt{I^2-J^2}}}\,
    \Big[\phi'-\alpha^2(a^2-l^2)^2\,\tau'\,\Big],\label{phi-trans-fin}
\end{align}
together with the transformation of~$x$ and~$r$ given by \eqref{def-x-and-r-1} and \eqref{def-x-and-r}.

Finally, the metric functions $P'$ and $Q'$ are explicitly obtained from the Astorino metric functions $\Delta_X$ and $\Delta_R$ using the relations \eqref{P-definition} and \eqref{Q-definition}, respectively. It turns out to be the quartics \eqref{P'Q'eqns}, where the parameters in the Pleba\'nski-Demia\'nski metric functions  are given by \eqref{direct_transformation_A-PD_parameters-simplified}.

Recall that throughout this Appendix we assumed ${a^2>l^2}$. The case ${a^2<l^2}$ leads to similar expressions, but with ${\sqrt{a^2-l^2}}$ replaced by ${\sqrt{|a^2-l^2|}}$, and ${I^2-J^2}$ replaced by ${I^2+J^2}$. In particular, it generalizes the definition of the parameters $J$ and $K$ in \eqref{defI-and -defJ-repeated} to \eqref{defI-and -defJ}, and yields a more general transformation \eqref{direct_transformation_A-PD-r}.

\end{document}